\documentclass{article}

\usepackage{arxiv}

\usepackage[utf8]{inputenc} 
\usepackage[T1]{fontenc}    
\usepackage{hyperref}       
\usepackage{url}            
\usepackage{booktabs}       
\usepackage{amsfonts}       
\usepackage{nicefrac}       
\usepackage{microtype}      
\usepackage{graphicx}
\usepackage{doi}

\usepackage{tikz}
\usepackage{amsmath}
\usepackage{cite}
\usepackage{amsmath,amssymb,amsfonts,amsthm}
\usepackage{algorithmic}
\usepackage{graphicx}
\usepackage{textcomp}
\usepackage{multirow}
\usepackage{tabularx}
\usepackage{threeparttable}
\usepackage{booktabs}
\usepackage{xcolor}
\usepackage{floatrow}
\usepackage{url}
\usepackage{floatflt}
\usepackage{subcaption}
\usepackage[inline,shortlabels]{enumitem}

\title{What Are the Chances?\\Explaining the Epsilon Parameter in\\ Differential Privacy}

\author{Priyanka Nanayakkara\thanks{The author conducted part of this work while visiting Columbia University.}\hspace{1.5mm}\thanks{The author conducted part of this work while visiting the Simons Institute for the Theory of Computing at UC Berkeley.}\\
        Northwestern University\\
        \texttt{priyankan@u.northwestern.edu}\\
	\And
	Mary Anne Smart\footnotemark[1] \\
	University of California San Diego \\
	\texttt{msmart@ucsd.edu} \\
	\AND
	Rachel Cummings\footnotemark[2]\hspace{1.5mm}\thanks{The author contributed equally to advising this work.} \\
	Columbia University \\
	\texttt{rac2239@columbia.edu} \\
	\And
	Gabriel Kaptchuk\footnotemark[3] \\
	Boston University \\
	\texttt{kaptchuk@bu.edu} \\
 	\And
	Elissa Redmiles\footnotemark[3]  \\
	Max Planck Institute for Software Systems \\
	\texttt{eredmiles@gmail.com} \\
}

\renewcommand{\headeright}{}
\renewcommand{\undertitle}{}
\renewcommand{\shorttitle}{What Are the Chances? Explaining the Epsilon Parameter in Differential Privacy}

\hypersetup{
pdftitle={What are the Chances?},
pdfsubject={differential privacy, usable security and privacy, privacy budgets},
pdfauthor={Priyanka Nanayakkara, Mary Anne Smart, Rachel Cummings, Gabriel Kaptchuk, Elissa Redmiles},
pdfkeywords={differential privacy, usable security and privacy, privacy budgets},
}

\newcommand{\samp}{\textsc{sample reports}}
\newcommand{\ratiotext}{\textsc{odds-text}}
\newcommand{\ratiovis}{\textsc{odds-vis}}
\newcommand{\no}{\textsc{no}}
\newcommand{\yes}{\textsc{yes}}
\newcommand{\true}{\textsc{true}}
\newcommand{\false}{\textsc{false}}

\urldef\myurl\url{https://osf.io/w59fv/?view_only=c42a3d68bf9d4f35abe488aab831e775}

\newcommand{\linReg}[4]{$\beta=#1$, $95\%\text{ CI}=[#2,#3]$}
\newcommand{\linRegFirst}[4]{$\beta=#1$, $95\%\text{ CI}=[#2,#3]$}
\newcommand{\ordRegFirst}[4]{$\text{OR}=#1$, $95\%\text{ CI}=[#2,#3]$}
\newcommand{\ordReg}[4]{$\text{OR}=#1$, $95\%\text{ CI}=[#2,#3]$}
\newcommand{\ordRegHalf}[3]{$95\%\text{ CI}=[#1,#2]$}

\begin{document}
\maketitle

\begin{abstract}
Differential privacy (DP) is a mathematical privacy notion increasingly deployed across government and industry. With DP, privacy protections are probabilistic: they are bounded by the privacy budget parameter, $\epsilon$. Prior work in health and computational science finds that people struggle to reason about probabilistic risks. Yet, communicating the implications of $\epsilon$ to people contributing their data is vital to avoiding privacy theater---presenting meaningless privacy protection as meaningful---and empowering more informed data-sharing decisions. Drawing on best practices in risk communication and usability, we develop three methods to convey
probabilistic DP guarantees to end users: two that communicate odds and one offering concrete examples of DP outputs.\\

We quantitatively evaluate these explanation methods in a vignette survey study ($n=963$) via three metrics: objective risk comprehension, subjective privacy understanding of DP guarantees, and self-efficacy. We find that odds-based explanation methods are more effective than (1) output-based methods and (2) state-of-the-art approaches that gloss over information about $\epsilon$. Further, when offered information about $\epsilon$, respondents are more willing to share their data than when presented with a state-of-the-art DP explanation; this willingness to share is sensitive to $\epsilon$ values: as privacy protections weaken, respondents are less likely to share data.
\end{abstract}


\section{Introduction}
Differential privacy (DP)~\cite{TCC:DMNS06} is a formal definition of privacy that has been integrated into several high-profile data analysis pipelines, including the 2020 U.S. Census data products~\cite{abowd2018us} and internal metric measurement tools at, e.g., Google~\cite{CCS:ErlPihKor14}, Apple~\cite{apple2017}, Microsoft~\cite{ding2017collecting}, and Uber~\cite{tezapsidis2017uber}.

As DP is increasingly applied to protecting people's privacy, it is vital that organizations deploying DP effectively communicate the privacy implications of implementation details that govern the strength of systems' privacy protections. Without such transparency, organizations risk engaging in ``privacy theater,''~\cite{smart2022,dwork2019differential,soghoian2011end} which may result in people falsely believing they are well-protected~\cite{tang2017privacy,CCS:CumKapRed21}.

While DP offers a precise framework for measuring worst-case privacy loss, research has found that non-experts struggle to form accurate assessments of the real-world privacy protections DP affords~\cite{CCS:CumKapRed21,SP:XWLJ20}. One source of confusion is the probabilistic (i.e., non-absolute) nature of DP's privacy protection. In particular, DP bounds privacy loss as a function of the unitless privacy budget parameter $\epsilon$. Differentially-private algorithms inject a calibrated amount of statistical noise inversely proportional to $\epsilon$ into either the data or analysis outputs (depending on the DP model), meaning higher values of $\epsilon$ correspond to weaker privacy protections.

Explaining probabilistic systems to end users (i.e., people contributing their data) is a challenging task, as observed by prior social science research on health risk communication~\cite{slovic2000perception,kahneman1982judgment}. Explaining probabilistic privacy risk, such as that created by DP, is a similarly---or perhaps an even more---challenging problem, given that the probabilistic nature of the system arises from the use of a complex, explicitly mathematical process, rather than variation in population-level behaviors.  Moreover, the privacy protections offered by differentially private mechanisms lack context, i.e., they are agnostic to the social context of a dataset or analysis. Privacy scholars, however, have theorized that people understand privacy contextually~\cite{nissenbaum2009privacy}. 

Despite the critical importance of $\epsilon$, many deployed DP systems only describe $\epsilon$ in technical documentation, while information about the privacy protection accessible to the general public glosses over 
the implications of the chosen $\epsilon$ altogether~\cite{CCS:CumKapRed21,dwork2019differential}. This is particularly problematic, as the values of $\epsilon$ used in practice, and thus the real-world privacy protections afforded by DP systems, vary wildly~\cite{censusparameters,desfontainesblog20211001}.

Prior research on explaining DP has either sidestepped the complicated task of explaining $\epsilon$ to end users~\cite{SP:XWLJ20,CCS:CumKapRed21,franzen2022private,karegar2022exploring} or focused on addressing the impact of $\epsilon$ for very specific deployments of DP~\cite{bullek2017,smart2022,elasticity}, e.g., explaining the randomized response mechanism~\cite{bullek2017}. As a result, we currently lack explanations of DP that include $\epsilon$ and can be used with all differentially-private mechanisms. Without access to such explanations, organizations deploying DP systems must either write new deployment-specific descriptions of DP that are unlikely to be scientifically evaluated, or risk leaving their users unable to make well-informed decisions.\looseness=-1

We fill this gap by developing and evaluating explanation methods for DP that \emph{directly} address the implications of $\epsilon$. The result of our work is a \textit{framework} for conveying $\epsilon$ to end users that is highly \emph{portable}, in that it can be adapted to  many deployment settings.  Our explanation methods avoid relying on describing the mathematical details of the mechanism and focus on the concrete ramifications of the choices a user might potentially face, e.g., if they are to share their data. This is a conceptual departure from prior work on DP communication, which focuses on the implications of using DP instead of running the equivalent, non-private system~\cite{smart2022,bullek2017} or learning attributes using information that is available even if an individual chooses not to share their data~\cite{franzen2022private,wood2018differential,SP:XWLJ20}. We also evaluate our methods by testing an instantiation of our explanation methods in a scenario with binary count queries. In Section~\ref{sec:discussion}, we offer direction for how our methods can be adapted to accommodate other queries and more complex scenarios.

\medskip
\noindent\textbf{New Explanations for Differentially-Private Systems.}
We draw on best practices in risk communication and usability~\cite{edwards2002explaining,lipkus2007numeric,gigerenzer1995improve,hoffrage1998using,galesic2009using,slovic2000perception,harbach2014using,kaptchuk2020good} to develop explanation methods designed to allow people to quickly and easily reason about probabilistic privacy guarantees under DP.

Specifically, we design three explanation methods for $\epsilon$. Our first explanation method (\ratiotext{}) leverages best practice methodology for risk communication to give a textual description of the odds that an information leak might occur if a person decides to share their data; this is a stylized version of the ``textbook'' understanding of DP~\cite{dwork2014algorithmic} which compares the outputs of differentially-private mechanisms applied to neighboring databases, each corresponding to a situation where a person does or does not share their data. Our second explanation method (\ratiovis{}) conveys the same information using a frequency-framed visualization approach which may help people with low numeracy skills more accurately make probability judgments~\cite{galesic2009using}. Our third explanation method (\samp{}) draws on prior work in usable security and privacy (S\&P) on improving user comprehension of S\&P technologies~\cite{harbach2014using} to provide people with several potential \emph{outputs} of the DP mechanism in an effort to make the implications of their data-sharing choice concrete. \looseness=-1

We evaluate the efficacy of these explanation methods using three metrics: (1) objective risk comprehension, (2) subjective privacy understanding, and (3) self-efficacy (personal belief in decision-making capacity~\cite{maddux2012self}). We additionally study the relationship between our explanation methods and (4) willingness to share data.

In summary, we are interested in answering the following research questions (RQs):

\begin{description}
\item[RQ1:] Which practices in risk communication work best for communicating the probabilistic privacy guarantees offered by DP?  Specifically, which practices are effective at increasing people's

        (a) \emph{objective risk comprehension} of DP guarantees, 

        (b) \emph{subjective privacy understanding} of DP guarantees, 

        (c) \emph{self-efficacy} around making data-sharing decisions?

\item[RQ2:] How do the explanation methods we developed influence people's data-sharing decisions? 
\end{description}

We answer our RQs via a vignette survey study ($n=963$) in which we embed our DP explanations into concrete information-sharing scenarios and evaluate them using the aforementioned criteria against each other and multiple control explanations.

\medskip
\noindent\textbf{Summary of Findings.}
We find that people have better objective risk comprehension (RQ1a) of DP protections when presented with odds-based explanations (\ratiotext{} or \ratiovis{}) than with \samp{}, which presents sample outputs from the privacy mechanism.

Despite our findings about objective risk comprehension, none of our explanations meaningfully improve people's \emph{subjective privacy understanding} (RQ1b), i.e., people \emph{feeling} as though they understand the privacy protection.

Further, to assess self-efficacy (RQ1c), we ask respondents if they feel as though they (1) have enough information to make a data-sharing decision and (2) are confident making said decision. The odds-based explanations we test increase people's sense that they have enough information to make a data-sharing compared to a state-of-the-art~\cite{SP:XWLJ20} explanation that does not feature information about $\epsilon$, suggesting that there is merit to not glossing over the probabilistic nature of DP privacy protection. However, our \samp{} method for explaining $\epsilon$ had the opposite effect: it reduced feelings of having enough information to make data-sharing decisions as compared to a very simple and clear description of the scenario without any probabilistic privacy protections, suggesting that this explanation actively confused respondents.

Interestingly, we do not find evidence that any of our explanations meaningfully impact people's confidence in making data-sharing decisions compared to a state-of-the-art explanation. Instead, we find that their overall concern about the ramifications of their data-sharing significantly relates to their confidence.

Last, we study the influence of our explanations on people's willingness to share data (RQ2). Our findings indicate that people are much more likely to share their data when presented with an explanation of DP that offers information about $\epsilon$ (compared to one that does not), regardless of which explanation method is used. Finally, when offered information about $\epsilon$, respondents' data-sharing decisions are sensitive to changes in $\epsilon$, empirically validating theoretical proposals that willingness-to-share depends on the strength of the privacy mechanism~\cite{dwork2014algorithmic}. \looseness=-1
\vspace{-2mm}
\section{Background}

\smallskip
\noindent
\textbf{Differential Privacy.}
\label{sec:background:DP}
DP \cite{TCC:DMNS06} is a mathematical privacy definition which ensures, at a high level, that the results of an analysis should be similar regardless of the inclusion of any given individual's information in that analysis. Differentially private mechanisms add carefully calibrated random noise at some point in the data analysis process in order to obscure details at the individual level while maintaining accuracy at the aggregate level. If too much noise is added, it will overwhelm the signal in the data, and the analysis results will be useless. If too little noise is added, the privacy protection offered to individuals may not be meaningful. The privacy budget parameter, $\epsilon$, controls this trade-off; a \textit{smaller} privacy budget provides a \textit{stronger} privacy guarantee. We state the formal definition of DP below:\looseness=-1

\emph{Definition: }[Dwork et al. \cite{TCC:DMNS06}]
A randomized algorithm $\mathcal{A}: \mathcal{D} \to \mathcal{R}$ is $\epsilon$-DP if for every pair of databases $D, D' \in \mathcal{D}$ that differ in at most one entry and for every subset $S \subseteq \mathcal{R}$, 
${\displaystyle Pr[\mathcal{A}(D) \in S] \leq e^{\epsilon}\cdot Pr[\mathcal{A}(D') \in S]}$.

Implementations of DP typically adopt either the \textit{local} or \textit{central} model.\footnote{Although, other models also exist~\cite{desfontaines2020sok}.} In the central model, a trusted curator stores the collected data and adds noise as necessary when releasing statistics, charts, or other aggregate insights about the data. In the local model, noise is added to each individual's data before it is sent to the curator.  While prior work has already explored the task of explaining the privacy budget to end users in the local model~\cite{bullek2017, smart2022,elasticity}, our study addresses the more challenging task of explaining the privacy budget in the context of the central model. 

One of the simplest mechanisms for achieving DP is the \textit{Laplace Mechanism}~\cite{TCC:DMNS06,dwork2014algorithmic}. In this mechanism, a data collector releases results of a simple counting query by adding noise sampled from a Laplace distribution centered at zero with scale parameter $\frac{1}{\epsilon}$. The resulting outcome is distributed according to $Lap(\mu, b=\frac{1}{\epsilon})$, where $\mu$ is the true value of the counting query before noise is added.\footnote{Note that the standard presentation of the Laplace Mechanism focuses on the distribution of the \emph{noise} rather than the \emph{output}. Our presentation is mathematically equivalent and is consistent with our use of the Laplace Mechanism in Section~\ref{explanation_methods:calculating_values}.}

\smallskip
\noindent
\textbf{Communicating DP to End Users.}
Prior work has begun to study the task of communicating DP to the general population~\cite{smart2022,CCS:CumKapRed21,bullek2017,SP:XWLJ20,karegar2022exploring,franzen2022private,xiong2022using, elasticity}. For the local model, Smart et al.~\cite{smart2022} and Bullek et al.~\cite{bullek2017} have explained the strength of privacy protections in terms of the probability of bits being ``flipped.'' However, this style of explanation does not work for the central model since noise is added at the aggregate level instead of at the individual level. Metaphors provide a different approach for explaining DP. For example, Karegar et al.~\cite{karegar2022exploring} use the metaphor of blurring images as an analogy for adding statistical noise to collected data. Such metaphors can help people understand that there exists a privacy-accuracy trade-off, such that increasing the injected noise strengthens the privacy guarantee but harms accuracy. However, these metaphors do not explain the implications of particular settings of $\epsilon$, a challenge we address in our work. 

Xiong et al.~\cite{xiong2022using} studied how to communicate the implications of the privacy budget for both privacy and accuracy of location data. The authors develop illustrations for the local, central, and shuffle models (see~\cite{bittau2017prochlo}) that show how the amount of added noise affects privacy and accuracy. They use heatmaps to compare the accuracy of collected location data before and after noise is added. They inject positive rather than unbiased noise to avoid the problem of negative counts that may confuse people who are unfamiliar with DP. In our study, we instead choose to embrace the sometimes unintuitive results produced by adding unbiased noise and investigate end users' perceptions of them.

Franzen et al.~\cite{franzen2022private} borrowed from the literature on quantitative risk communication to explain the protections offered by DP. Although quantitative risk communication formats can aid comprehension, individuals with low numeracy skills struggle to understand these explanations. We include a measure of numeracy skill in our survey to determine whether our explanations similarly disadvantage individuals with low numeracy skills. An important difference between Franzen et al.~\cite{franzen2022private} and our work is the comparison probability (to the probability of a negative outcome given that an individual shares their data) we each present: Franzen et al. present the probability of a negative outcome given no data collection takes place, while we show the probability of a negative outcome given the individual does not share their data, but all other factors remain the same. Because people rarely have the power to immediately stop an entire data collection process, we suggest that it is important to explore this separate decision context in an effort to closely align with real-world decisions people make.

\smallskip
\noindent
\textbf{Supporting Decision-Making Around $\epsilon$.}
\label{sec:relatedwork:decisionmaking}
Research on communicating implications of $\epsilon$ has tended to focus on data curators or analysts who are setting privacy budgets. For example, there have been several interfaces for DP (DPComp~\cite{hay2016exploring}, Overlook~\cite{thaker2020overlook}, PSI ($\Psi$)~\cite{gaboardi2016psi}, Bittner et al.~\cite{bittner2020understanding}, DPP~\cite{john2021decision}, ViP~\cite{nanayakkara2022visualizing}) that portray accuracy and/or risk implications of $\epsilon$ to support more informed privacy budget setting. Although these tools are aimed at data analysts and curators, they are also relevant to communicating DP to non-experts because these analysts/curators typically are not assumed to have DP background or expertise. As such, these tools must express relevant DP concepts well enough to support decision-making about privacy budgets. At the same time, we note that end users are usually not tasked with setting or allocating privacy budget, but rather must make individual data-sharing decisions.

Hsu et al.~\cite{hsu2014differential} propose an economic framework for people considering sharing their data, e.g., as part of a scientific study, to weigh monetary costs of sharing versus not sharing their data. Wood et al.~\cite{wood2018differential}, in explanations of $\epsilon$ in a primer on DP, similarly frame data-sharing decisions in terms of worst-case monetary losses (e.g., in terms of increases to insurance premiums) people could incur if they share their data under DP. Heffetz and Ligett~\cite{heffetz2014privacy} describe $\epsilon$ to economists in the context of calculating a mean salary value, focusing primarily on accuracy outcomes. Finally, Lee and Clifton~\cite{lee2011much} model disclosure risk by considering a potential attacker who conducts a Bayesian update on their beliefs of whose information is included in an analysis based on seeing a release from a differentially-private mechanism. Our odds-based explanation methods similarly model an attacker's updated beliefs given a DP output. 

\smallskip
\noindent
\textbf{Probabilistic Risk Communication.}
Many studies have identified best practices for effective probabilistic risk communication, especially in the medical context~\cite{edwards2002explaining,lipkus2007numeric}. Prior work has found benefits of framing probabilities as frequencies~\cite{gigerenzer1995improve,hoffrage1998using}. One challenge in probabilistic risk communication is that people often misinterpret probabilities expressed as ratios---for example, people may mistakenly interpret an event with a probability of 1 out of 10 as less likely than an event with a probability of 10 out of 100, simply because the former ratio is expressed with smaller numbers~\cite{alonso2003irrational,yamagishi199712}. Thus, it is best practice to use a consistent denominator when presenting ratios for comparison~\cite{lipkus2007numeric}. Frequency-framed visualizations, such as icon arrays, can also complement numeric risk communication. Compared to purely numeric presentations of risk, icon arrays may improve understanding particularly among people with low-numeracy skills~\cite{galesic2009using}. We incorporate these findings into our explanations of DP by framing probabilities as frequencies and employing icon arrays.

\section{Explanation Methods for $\epsilon$}
\label{sec:methods:explanations}
We introduce three methods to explain $\epsilon$ to end users. These methods work for two common data-sharing settings: one where providing data is \textit{optional}, so people must decide whether to participate (or opt-out), and one where providing data is \textit{mandatory}, so people must decide whether to respond truthfully (or respond untruthfully).

\begin{figure*}[ht]
\RawFloats
	\centering
	\begin{minipage}[t]{.29\columnwidth}
		\includegraphics[width=0.98\textwidth]{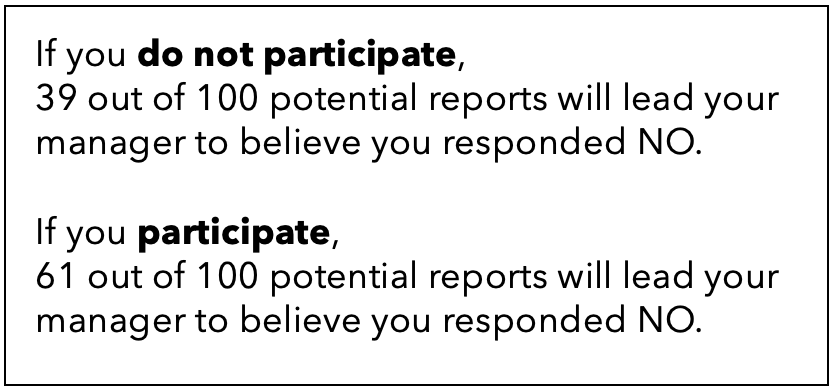}
		\subcaption[t]{\ratiotext}
		\label{fig:ratiotext}
	\end{minipage}%
	\begin{minipage}[t]{.29\columnwidth}
		\includegraphics[width=0.98\textwidth]{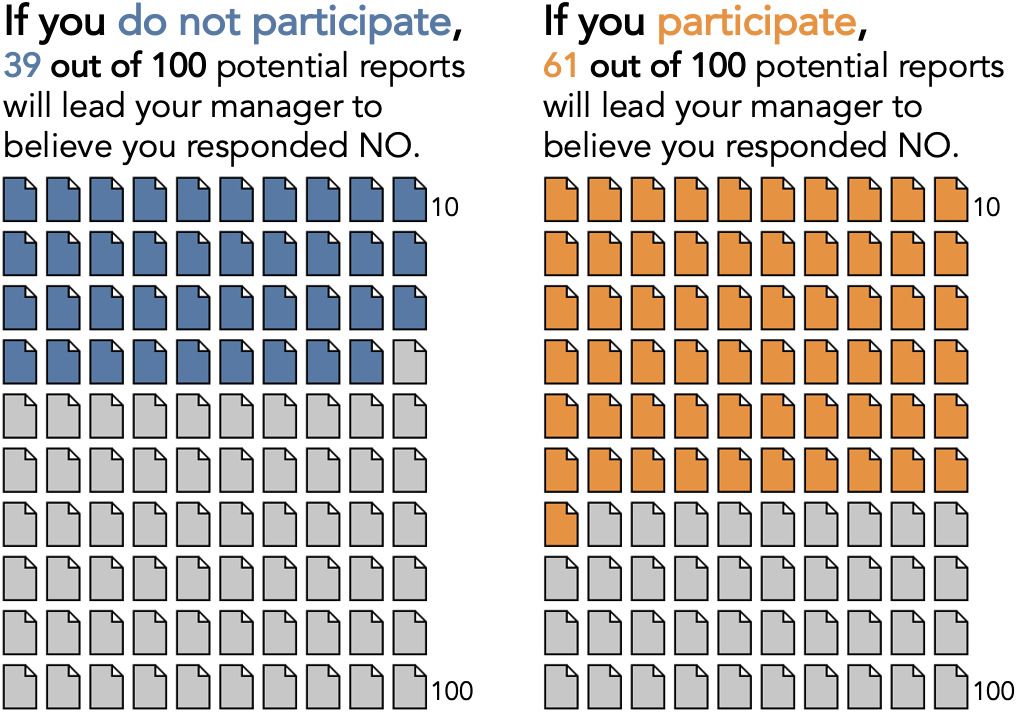}
		\subcaption{\ratiovis}
		\label{fig:ratiovis}
	\end{minipage}
 	\begin{minipage}[t]{.4\columnwidth}
		\includegraphics[width=0.98\textwidth]{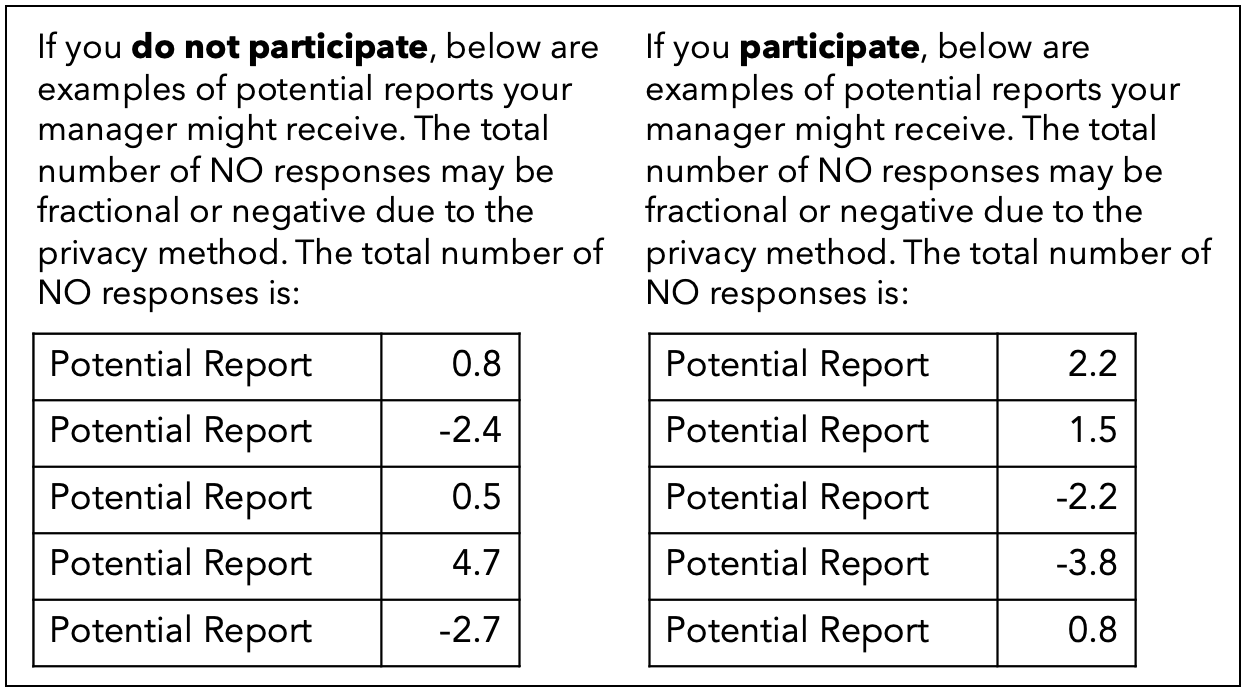}
		\subcaption{\samp}
		\label{fig:samp}
	\end{minipage}
	\caption{\footnotesize Examples of our explanation methods applied to the manager's performance review scenario under the optional setting and $\epsilon=0.5$.}
 \label{fig:examples}
\end{figure*}

Drawing on best practices from the literature in health risk communication~\cite{edwards2002explaining,lipkus2007numeric,hoffrage1998using}, we develop two explanation methods (\ratiotext{} and \ratiovis) that focus on explaining the \textit{odds} of a negative event occurring by contextualizing privacy guarantees in terms of outcomes that could occur based on decisions users can make. We develop a third method (\samp) that provides concrete examples of the protected data, based on findings that indicate concrete examples help people comprehend S\&P topics~\cite{harbach2014using}. Examples of each explanation method instantiated in our survey scenario (see Section~\ref{sec:manager_scenario}) are in Figure~\ref{fig:examples}.

\subsection{Description Approaches}\label{s.methodsdesc}
\textbf{\ratiotext{.}}
In line with research~\cite{slovic2000perception} finding that people reason more effectively about the odds of a risk when framed as frequencies versus percentages, including in the context of privacy decisions~\cite{kaptchuk2020good} and DP specifically~\cite{franzen2022private}, we present all probabilities in the \ratiotext{} explanation as frequencies (Figure~\ref{fig:ratiotext}). Specifically, probabilities are shown in the form of ``$z$ out of $100$ potential DP outputs'' where $z$ is a natural number and ``DP outputs'' can be customized to specific scenarios (e.g., if the DP output is published in a report, the explanation may instead say ``potential reports''). Specifically, this explanation method comprises of two probabilities corresponding to the chances that an adversary $\mathcal{A}$ believes, based on prior knowledge combined with a DP output, that a data sharer provided information corresponding to the actual value $d_{\text{true}}$ if the data sharer participates vs. does not participate OR responds truthfully vs. untruthfully. Data sharers have immediate agency over these actions, and hence showing probabilities aligning with these actions is directly relevant to their data-sharing decision-making process. Specifically, \ratiotext{} explanations take the following form:
\begin{quote}
\emph{
    If you [do not participate/respond $d_{\text{false}}$], $x$ out of $100$ potential [DP outputs] will lead $\mathcal{A}$ to believe you responded $d_{\text{true}}$.\\
    If you [participate/respond $d_{\text{true}}$], $y$ out of $100$ potential [DP outputs] will lead $\mathcal{A}$ to believe you responded $d_{\text{true}}$.
    }
\end{quote}

\smallskip
\textbf{\ratiovis{}.}
Research has found that icon arrays---a frequency-framed visualization approach (Figure~\ref{fig:ratiovis})---sometimes help people with low-numeracy skills more accurately estimate risk reduction~\cite{galesic2009using}. Thus, we add an icon array to the text-based description of the risk ratios in the \ratiotext{} condition. We use icon arrays to help people concretely visualize that there are many potential DP outputs, but $\mathcal{A}$ will only receive one. The shape of the icon can be adapted to suit the scenario. We fill in icons top to bottom, as this arrangement has been shown to be optimal for supporting accurate probability judgments~\cite{xiong2022investigating}. Icon colors follow from the Tableau 10 color palette~\cite{tableau10}, which was designed keeping in mind common forms of color-vision deficiencies.

\smallskip
\textbf{\samp{}.}
Drawing on prior work in S\&P showing that concrete examples improve user comprehension of privacy enhancing technologies and secure behavior~\cite{harbach2014using}, the \samp{} method shows five potential DP outputs from an analysis if the data sharer were to participate/respond truthfully (i.e., share $d_{\text{true}}$) and five potential DP outputs if they were to not participate/respond untruthfully (i.e., provide no data or share $d_{\text{false}}$) (Figure~\ref{fig:samp}). Presenting both sets of potential outputs allows the data sharer to make comparisons between how these values would differ based on their decision, and the extent to which their survey response sways the DP output. 

\subsection{Contextualizing Explanations}
\label{sec:methods:computing}
In order to leverage the description approaches outlined above, we require the following: (1) a concrete scenario in which the data sharer should think about the explanation and (2) the values for parameters described above (e.g., probabilities for both odds-based methods). In this section, we introduce a hypothetical data-sharing scenario and outline how to appropriately compute values for explanations for the given scenario. We use this hypothetical scenario in surveys with respondents (more details in Section~\ref{sec:evaluationmethods}). As detailed below and in Section~\ref{sec:discussion}, our methods and calculation techniques for each method can be extended to numerous other DP applications.

\subsubsection{Workplace Evaluation Scenario}
\label{sec:manager_scenario}

Imagine that an employee (the data sharer) is asked to share an evaluation of their manager ($\mathcal{A}$), but fears their manager will retaliate against them if they believe the employee reviewed them negatively. Everyone reporting to the manager is asked to share an evaluation, which specifically asks the following \yes{}/\no{} question:  \emph{Do you feel adequately supported by your manager?} This scenario describes data collection and analysis occurring under the central DP model: the company will collect un-noised answers and create a report with the total number of \no{} responses calculated using the Laplace Mechanism with a particular $\epsilon$. The report will \emph{not} include names of team members and how they responded, however. Note that our explanation methods can be used to convey privacy strength of various other DP mechanisms (e.g., Gaussian) simply by plugging in other noise distributions in the analysis that follows.

 Since DP guarantees are often framed through a worst-case lens, we further design the scenario to represent a worst-case situation where the manager has prior knowledge on how all the other teammates will respond. For concreteness, we suppose that they will all respond \yes{}, while the employee wants to respond \no{} ($d_\text{true}$), thus putting the employee at risk of being singled out in summary statistics. The other teammates' responses can be modified to suit other contexts, resulting in modifications to $\mu$ in the following analysis. We also imagine that the manager's prior belief that the employee will respond \no{} is 50\%. See Appendix~\ref{appendix:priors} for how the analysis specifically for \ratiotext{} and \ratiovis{} can be updated to accommodate non-uniform priors.

\subsubsection{Calculating Values for Explaining $\epsilon$}
\label{explanation_methods:calculating_values}
We describe how to calculate values for explanations in the workplace data-sharing context described above. Values computed for our explanation methods are the same regardless of whether the employee is in the mandatory or optional data-sharing setting.\footnote{These values are the same in our case because we consider count queries, and hence the sensitivies do not change.}

\smallskip
\textbf{\ratiotext{} and \ratiovis{}.}
We calculate values for the \ratiotext{} and \ratiovis{} explanations by modeling the manager's guessing process. We assume that the manager correctly believes all other teammates will respond \yes{}, consistent with the scenario details described in Section~\ref{sec:manager_scenario}. Thus, the true count of \no{} responses is either $0$ or $1$, depending on whether the employee participates/responds truthfully. Values consistent with this scenario for multiple privacy budgets are shown in Table~\ref{table:ratiotextvalues}.

Specifically, we model the manager's inference process as estimating the employee's most likely survey response, given the observed DP output. Using distributional knowledge of the DP mechanism, the manager compares posterior probabilities of the employee's response given the DP output to determine a maximum likelihood estimate (MLE). Let the output of the DP mechanism be $r$. The manager compares $\Pr[E = r \;|\; E \sim Lap(\mu=1, b =  \frac{1}{\epsilon})]$ with $\Pr[E = r \;|\; E \sim Lap(\mu=0, b =  \frac{1}{\epsilon})]$, respectively corresponding to when the employee reports truthfully/participates and when they lie/do not participate, and then guesses the action corresponding to a higher posterior probability. We compute a threshold value $r_\text{threshold}$ where $\Pr[E = r_\text{threshold} \;|\; E \sim Lap(\mu=1, b = \frac{1}{\epsilon})]=\Pr[E = r_\text{threshold} \;|\; E \sim Lap(\mu=0, b = \frac{1}{\epsilon})]$ and report $\Pr[r<r_\text{threshold}]$ (i.e., $\frac{x}{100}$ in Table~\ref{table:ratiotextvalues}) and $\Pr[r>r_\text{threshold}]$ (i.e., $\frac{y}{100}$ in Table~\ref{table:ratiotextvalues}). \looseness=-1

\begin{table}[t]
\begin{tabular}{l|l|l|l}
$\epsilon=0.1$                                          & $\epsilon=0.5$                                          & $\epsilon=2$                                            & $\epsilon=4$                                           \\ \hline \hline
\begin{tabular}[c]{@{}l@{}}$x=48$\\ $y=52$\end{tabular} & \begin{tabular}[c]{@{}l@{}}$x=39$\\ $y=61$\end{tabular} & \begin{tabular}[c]{@{}l@{}}$x=18$\\ $y=82$\end{tabular} & \begin{tabular}[c]{@{}l@{}}$x=7$\\ $y=93$\end{tabular}
\end{tabular}
\caption{ \footnotesize Values for odds-based explanation methods.}
\label{table:ratiotextvalues}
\end{table}

\smallskip
\textbf{\samp{}.}
In our scenario, the manager will receive a report with the total number of \no{} responses. To obtain potential outputs for \samp{} under a given $\epsilon$, we make five random draws from each of $Lap(\mu=1,b=\frac{1}{\epsilon})$, corresponding to potential DP outputs when $d_\text{true}$ is shared, and  $Lap(\mu=0,b=\frac{1}{\epsilon})$, corresponding to potential DP outputs when there is no participation/$d_\text{false}$ is shared. We do not post-process sampled values, meaning that they can be fractional or negative. As such, we include a statement preceding these values explaining: \textit{``The total number of \no{} responses may be fractional or negative due to the privacy method.''}

\section{Evaluation}
\label{sec:evaluationmethods}
We evaluate our explanation methods by conducting online vignette surveys ($n=963$), which are designed to mimic real-world decision-making behavior~\cite{hainmueller2015validating}, where we present respondents with the survey scenario described in Section~\ref{sec:manager_scenario} and an explanation of $\epsilon$ created using one of our three methods, as detailed in Section~\ref{s.methodsdesc}. Our institutions' IRBs approved this research.

\subsection{Survey Scenario}
Survey respondents are told to imagine themselves as the data sharer/employee described in Section~\ref{sec:manager_scenario}, either in the optional or mandatory setting. Respondents who see a scenario in the optional setting are told that while participation in their company's survey is optional, participating means they will necessarily respond truthfully (they will respond \no), i.e., if they participate they cannot lie. On the other hand, respondents who see the mandatory setting are explicitly told they can either respond truthfully (\no) or untruthfully (\yes). Across settings, scenario details are the same except for slight differences in wording expressing whether the company's survey is mandatory or optional.

Following best practices for vignette studies~\cite{riley2021vignettes}, we designed this scenario to mimic common performance evaluations conducted across workplace and educational settings, thus increasing the chances that respondents would find the scenario believable and relatable. To ensure that all respondents read our explanations of privacy guarantees with similar assumptions about why they would want their data protected in the hypothetical scenario, we clearly define the negative consequences that could transpire if the manager were to correctly guess a negative survey response (\textit{``Your manager may retaliate if they believe you responded \no. For example, they might give you a negative performance review, assign you extra work, or try to get you fired.''}). We then ask the respondent to imagine that the team's performance reviews will be protected using a ``privacy protection method'' and do not include the term ``differential privacy'' anywhere in the survey to prevent respondents from searching for external materials.

The impact of the privacy-protection mechanism is then explained using a method described in Section~\ref{sec:methods:explanations}, with one of four privacy budgets ($\epsilon\in\{0.1,0.5,2,4\}$), which represent a range of privacy protection strengths. There is no standard for setting $\epsilon$~\cite{dwork2019differential}, so we chose these values to represent what is often recommended in the academic DP literature (small $\epsilon$ values, like $0.1$) and larger values that are more consistent with real-world deployments~\cite{desfontainesblog20211001}. We note that some real-world applications of central DP use privacy budgets much larger than our largest value (e.g., the U.S. Census Bureau set a total privacy budget of $19.61$ for the 2020 Census redistricting data~\cite{censusparameters}), but these budgets refer to $\epsilon$ accumulated over many queries, whereas our scenario includes just one.  \looseness=-1

All respondents who receive \samp{} under the same $\epsilon$ are presented with the same set of random draws. That is, to maintain consistency in what respondents see, we make these draws in advance and do not dynamically make new draws for each respondent. All generated values shown in \samp{} explanations are available in Appendix~\ref{appendix:surveyinstrument:samplereports}.\footnote{For reproducibility, we seeded the mechanism with the date. Our code for making draws is available here: \myurl.} Values shown in both odds-based explanations are in Table~\ref{table:ratiotextvalues}.

We tested our questionnaire via cognitive interviews ($n=12$), following best practices~\cite{redmiles2017summary} and using a think-aloud protocol~\cite{willis2005cognitive}, with potential study respondents to further refine the scenario for clarity and believability. Based on these interviews, we iterated on the introduction to our explanations and further specified potential negative consequences of information disclosure. We additionally tested our questionnaire via expert reviews ($n>10$) by experts in DP, survey methodology, and visualization.

\subsection{Experimental Design}
\label{sec:methods:experimentaldesign}
We use a $3\times4\times2$ between-subjects study design where each respondent sees one explanation, computed using a particular explanation method (\ratiotext, \ratiovis, \samp) and privacy budget ($\epsilon\in\{.1,.5,2,4\}$), in a given scenario type (optional, mandatory). We also have two control explanations to which we compare our experimental explanations: one where there are no privacy protections (\emph{deterministic control}) and another that includes a high-level explanation of DP that does not mention $\epsilon$, which is adapted from Xiong et al.~\cite{SP:XWLJ20} (\emph{Xiong et al. control}).\looseness=-1

\smallskip
\noindent
    \textbf{Deterministic Control:} The worst-case situation for a hypothetical employee in our vignette is that no privacy protection is applied. In this case, the risk of the negative consequence in the scenario is deterministic (i.e., respondents can expect deterministic outcomes in their manager's beliefs): If the respondent participates/answers \no{}, their manager will believe they responded \no{} with probability 1 and if they do the opposite, their manager will believe they responded \no{} with probability 0. Not only is this ``explanation'' a worst case, but prior work~\cite{slovic2000perception,bogardus1999perils,yamagishi199712,alonso2003irrational} on risk communication also suggests that it will be the simplest for respondents to understand (objective risk comprehension) and may give them the greatest self-efficacy because of its determinism, in contrast to our probabilistic explanations. Hence, we use this deterministic setting as a control: to do so, we include the same scenario text as in the experimental conditions but omit the stylized description of DP and explanation of privacy guarantees.

\smallskip
\noindent
    \textbf{Xiong et al. Control:} We also compare our experimental explanations to the current state-of-the-art explanation of DP from Xiong et al.~\cite{SP:XWLJ20}. 
    Because this explanation does not offer information about $\epsilon$, it helps us assess the impact of adding information about $\epsilon$ on people's understanding of DP protections, self-efficacy in data-sharing decisions, and willingness to share data. To present this control, we include the same scenario text as in the experimental conditions but replace our stylized DP description and privacy explanation with a DP description adapted from Xiong et al.~\cite{SP:XWLJ20} (precise wording in Appendix~\ref{sec:surveyinstrument:xiongbaseline}).
    While Xiong et al. propose several descriptions, we adapt their ``DP without names'' description since it aligns best with our scenario, and because their evaluation indicates that people found it easy-to-understand and that it supported comprehension on a relevant evaluation question about third-party viewers of the data. 

\subsubsection{Evaluation Metrics and Willingness to Share Data}
\label{sec:methods:experimentaldesign:eval}
Respondents answered questions to evaluate our explanations on three metrics: (1) objective risk comprehension, (2) subjective privacy understanding, and (3) self-efficacy. We also study the relationships between our explanations and (4) respondents' willingness to share data.

\smallskip
\noindent
    \textbf{(1) Objective Risk Comprehension:} We included two \true/\false{} questions to evaluate whether the explanations help people understand the risk inherent in the scenario: whether or not their manager will think they responded \no{}. In addition to options for \true{} and \false, we also provide an ``I don't know'' option~\cite{redmiles2017summary} to minimize random guesses that are correct by chance. Prior work on communicating DP to people~\cite{SP:XWLJ20,smart2022,franzen2022private} has similarly asked objective-risk-comprehension questions. 
    
    The first question, in the mandatory setting, reads: 
    \begin{quote}
        \textit{My manager is more likely to think I responded \no{} (i.e., respond truthfully) if I respond \no{} on the survey than if I respond \yes{} (i.e., respond untruthfully).}
    \end{quote}
    The version shown in the optional setting is nearly identical but asks about the manager's beliefs if the person were to participate/not participate.
    
    Respondents in all experimental conditions and the deterministic control answer this question. Respondents in the Xiong et al. control do not answer this question, as the explanation does not specify $\epsilon$; thus ground-truth answers do not exist.\footnote{Because the Xiong et al. control indicates that DP is used, it can be argued that there is a correct answer (\true{}) to the first objective-risk-comprehension question. 
    However, as $\epsilon$ tends toward zero, the difference in probabilities goes to zero. Hence for large $\epsilon$ the correct answer would clearly be \true{}, but for small $\epsilon$ values very close to 0, there may be little practical distinction between the probabilities and \false{} would be approximately correct.} For our experimental conditions, the correct answer under all privacy budgets we test is \true. For the deterministic control, the correct answer is \true.
    
    The second objective-risk-comprehension question is more challenging but takes a similar form to the first question (again we provide options of \true{}, \false{}, and ``I don't know''):
    \begin{quote}
    \textit{My manager is more than twice as likely to think I responded \no{} if I respond \no{} (i.e., respond truthfully) on the survey than if I respond \yes{} (i.e., respond untruthfully).}    
    \end{quote}
    For the same reason as the first objective-risk-comprehension question, only respondents in the experimental conditions and deterministic control see it. For our experimental conditions, the correct answer is \true{} for  $\epsilon\in\{2,4\}$ and \false{} for $\epsilon\in\{0.1,0.5\}$. For the deterministic control, the correct answer is \true. We create a score ranging from $0$--$2$, which is the total number of correctly-answered objective-risk-comprehension questions. ``I don't know'' responses are considered incorrect.
 
    \smallskip
\noindent
    \textbf{(2) Subjective Privacy Understanding:} We ask respondents to rate their confidence that they understand the privacy protection on a 4-point semantic scale (not at all confident--very confident). Previous studies explaining DP to people have similarly asked questions around subjective privacy understanding of DP~\cite{bullek2017,smart2022,franzen2022private}. Respondents are also asked to describe the privacy protection in their own words via an open-text response. Only respondents in the experimental conditions and Xiong et al. control see these questions; respondents who receive the deterministic control are not shown this question because this control does not describe privacy protections.

        \smallskip
\noindent \textbf{(3) Self-Efficacy:} To understand how empowered people feel to make data-sharing decisions based on our explanation types, respondents are asked three questions about confidence in their decision making. First, they are asked to rate on a 4-point semantic scale their confidence that they have enough information to decide which action to take (similar to a question Franzen et al.~\cite{franzen2022private} classify as ``Subjective Understanding''). Second, they are asked to describe in an open-text response what further information, if any, they would like to have to help them with their decision. Third, they are asked to rate on a 4-point semantic scale their confidence in deciding which action to take. These questions are shown to respondents in all conditions.

\smallskip
\noindent
    \textbf{(4) Willingness to Share Data:} To assess respondents' willingness to share data, each respondent was asked whether they would participate in the survey (in the optional condition) or answer truthfully on the survey (in the mandatory condition). 
    They were then asked to explain their reasoning in an open-text response. Although respondents could navigate backward through the survey at any time, their answer to this question was locked after advancing. All respondents were shown these questions.

The 4-point semantic scales were randomly reversed for roughly half of respondents in line with best practices~\cite{redmiles2017summary}. If scales were reversed for a particular respondent, all corresponding scales in their survey were also reversed.

\subsubsection{Questionnaire Structure}

Survey respondents are first instructed that they will read a fictional scenario and answer follow-up questions. Next, they read the first section of the scenario, which introduces the hypothetical survey about their manager, how they want to respond \no, and the potential repercussions of doing so.
We then assess whether respondents are indeed concerned about these consequences by asking them to rate their level of concern on a 4-point semantic scale (not at all concerned--very concerned), which we later refer to as ``baseline concern'' in Section~\ref{sec:results}. We also include an easy-to-answer comprehension check question, and filter out respondents who fail to answer this question correctly after two attempts.

Respondents in the deterministic control end the scenario at this point. Respondents in the Xiong et al. control read the adapted explanation of DP. Respondents in the experimental conditions 
are provided with the following abstraction of a random distribution:
\begin{quote}
\textit{Your company will not report exactly how many employees on your team responded \no. Instead, they will generate many potential reports by using a statistical method to modify the total number of \no{} responses. So, each potential report may show a number somewhat lower or higher than the actual number of \no{} responses. Only \underline{ONE} report will be randomly sent to your manager.}
\end{quote}

Then, the respondent is shown one computed explanation.

Subsequently, all respondents answer a series of questions on willingness to share data, objective risk comprehension, subjective privacy understanding, and self-efficacy. Finally, respondents answer questions on numeracy skills~\cite{lipkus2001general}, internet skills~\cite{hargittai2010digital}, and demographics (gender, age, race, education, computer science/IT educational/work background, and income). The final question is an open-text question for bot detection~\cite{kennedy2020assessing}. Respondents may go back to previous pages of the survey to review any information and are also provided with links to PDFs with complete descriptions of the scenario and privacy protections for easy access.
The full survey text is available in Appendix~\ref{appendix:surveyinstrument:scenariotext}.

\subsection{Participant Recruitment}
Participants were all recruited on Prolific, based in the U.S., and were at least 18 years old. We performed a power analysis to estimate the appropriate sample size for the survey~\cite{pwr}; 963 respondents completed the online survey and 12 completed cognitive interviews. We paid cognitive interview participants \$7.50 for 15 minute interviews (\$30/hour). Survey respondents were paid \$2.95 in each of the experimental conditions (median completion time: 10.9 minutes; $\sim$\$16/hour) and \$2.30 in each control condition (median completion time: 9.2 minutes; $\sim$\$15/hour). A detailed breakdown of survey respondents' demographics can be found in Appendix~\ref{app:demographics}.

\subsection{Analysis}
\label{sec:methods:analysis}
Our analysis\footnote{Code available here: \myurl.} aims to study (1) the effect of our explanation methods on several dependent variables (DVs) and (2) relationships between these outcomes and sociodemographic attributes, which prior work finds may influence privacy-related decisions (see, e.g., \cite{madden2017privacy,redmiles2017digital,hargittai2010facebook,park2015men,redmiles2018net,rainie2013anonymity, habib2018away,litt2013understanding}). To guide our analysis, we first construct causal directed acyclic graphs (DAGs) informed by prior work on relationships between demographic attributes and privacy concerns (e.g.,~\cite{o2001analysis}), internet skills (e.g.,~\cite{hargittai2019internet,hargittai2015mind}), and numeracy skills (e.g.,~\cite{galesic2010statistical}). These DAGs are in Appendix~\ref{appendix:DAG}.

For our primary analysis, we construct a set of regression models studying the effect of our independent variables (IVs)---explanation method (categorical, see details below), scenario setting (binary: optional or not), and privacy budget ($\epsilon$ as a numeric)\footnote{We include $\epsilon$ instead of $e^{\epsilon}$ in models because $e^{\epsilon}$ artificially compresses relevant privacy conditions, especially for small values of $\epsilon$~\cite{elasticity}.}---on our DVs: objective risk comprehension, subjective privacy understanding, self-efficacy, and willingness to share data. We construct logistic regression models for binary DVs (willingness to share data), ordinal regression models for ordinal DVs (subjective privacy understanding and both self-efficacy measures), and linear regression models for continuous DVs (number of correctly-answered objective-risk-comprehension questions). 

When constructing regression models, we treat the explanation IV as a categorical variable. Depending on the DV in the model, we use one or both of the control explanations as the reference (baseline) level for comparison. As described in Section~\ref{sec:methods:experimentaldesign:eval} not all DVs are applicable for both control conditions. In cases where both controls are applicable, we construct multiple models, with each control as the reference level for the explanation IV, respectively. In order to study relationships between $\epsilon$ and the IVs, we cannot use either control condition because there exists no state-of-the-art or control for presenting information about $\epsilon$. To compare the efficacy of our odds-based methods (\ratiotext{} and \ratiovis{}) with our concrete-example usability-based method (\samp{}), we also construct models for each DV using \samp{} as the explanation reference level. 

Next, we conduct a secondary analysis on relationships between our IVs and age (numeric), education (categorical), gender (binary\footnote{We provided respondents four options (man, woman, non-binary, self-describe) and allowed them to choose multiple~\cite{scheuerman2020hci}.
20 respondents identified as non-binary and one as fluid gender. Our sample sizes are too small to draw meaningful conclusions about how each of these groups interprets our explanations. Thus, for modeling purposes, we code gender as whether someone identifies as a man (regardless of their other gender identities).}), baseline concern (ordinal), internet skills (numeric), and numeracy skills (numeric), which prior work finds may influence respondents' ability to interpret numerically-related information such as that presented in our explanations~\cite{slovic2000perception,franzen2022private}. Note that we include baseline concern in this analysis in line with best practice guidance on vignette surveys and on privacy vignettes in particular~\cite{martin2016putting}. We fit models with only demographics (age, education, gender) and models with baseline concern, internet skills, and numeracy skills adjusted for said demographics.

To support results from the statistical analysis, we qualitatively analyze the open-text responses to help contextualize salient quantitative results. Two of the authors reviewed a subset of about $10\%$ of the respondents' open-text responses and together developed a codebook (available in Appendix~\ref{app:codebook}) capturing themes from responses that help illustrate findings from the statistical analysis. One of the authors then coded an additional subset of about 20\% of responses to ensure that we had captured a majority of general sentiments in our codebook; no additional codes were identified during this second round of coding. In total, we coded $283$ responses. Because the qualitative data are neither quantified via counts nor our primary research focus, we do not report inter-coder reliability~\cite{mcdonald2019reliability}.\looseness=-1

\subsection{Limitations}
Our study's results are limited by multiple factors which apply to large-scale survey studies. First, our sample may have failed to capture a representative population. Research has found that Prolific has relatively high external validity for questions about beliefs and perceptions related to privacy~\cite{tang2022replication}. However, as is typical for crowdsourced studies, our sample skews toward younger and more educated individuals, and thus does not fully represent the U.S. population. Our study is also conducted on people based in the U.S., which may limit the applicability of our findings to cultural contexts outside the U.S. Second, we aimed to make the survey scenario as realistic and understandable to respondents as possible by refining it through several cognitive interviews while maintaining a reasonable survey length. However, it is possible that in obscuring certain details, the scenario does not reflect important aspects of various workplace environments that may impact how people make decisions about sharing data. Third, although we aimed to elicit respondents' actual responses by following best practices~\cite{redmiles2017summary} such as providing ``I don't know'' answer choices where applicable, it is possible that respondents' answers do not always align with their actual feelings/decisions they would make in similar real-life scenarios. Prior work suggests that vignette studies can be powerful tools for understanding real-life behavior, especially when respondents are highly engaged. We suspect that the high level of concern expressed by respondents about the fictional scenario---over 75\% said they were ``concerned'' or ``very concerned''---suggests a high level of engagement. Finally, our survey only focused on the data-sharing scenario of workplace surveys; our findings may not generalize directly to other settings. Although we examine our explanations within a single scenario, we note that the explanation methods we develop for DP are scenario-agnostic, so future work could port these explanation methods into new scenarios as needed.

\section{Results}
\label{sec:results}
Based on our survey results, we seek to evaluate the effectiveness of our explanations (\textbf{RQ1}) and study how our explanations impact people's data-sharing decisions (\textbf{RQ2}).

\subsection{Effectiveness of Explanations (RQ1)}
We evaluate the effectiveness of our three explanations, \ratiotext{}, \ratiovis{}, and \samp{}, via three metrics: (1) objective risk comprehension of probabilistic DP guarantees, (2) subjective privacy understanding of DP guarantees, and (3) self-efficacy. 

\subsubsection{Objective Risk Comprehension (RQ1a)}
\label{sec:results:objcomp}
We construct a linear regression model (Table \ref{tab:objperceivedcomprehension}, left)\footnote{Tables that include $p$-values are available here: \myurl.} where the DV is the total number of correctly-answered objective-risk-comprehension questions (per respondent) following the methods in Section~\ref{sec:methods:analysis}.
We find that objective comprehension of privacy risks among respondents shown \ratiovis{} explanations was higher than that of respondents shown the deterministic control (\linRegFirst{0.28}{0.10}{0.46}{}). On the other hand, respondents shown \samp{} performed significantly worse on objective-comprehension questions compared to respondents shown the deterministic control (\linReg{-0.43}{-0.61}{-0.25}{x}).

We also construct a linear regression model with the same DV, but where the explanation reference level is the \samp{} explanation method, which allows us to include $\epsilon$ in the model (as described in Section~\ref{sec:methods:analysis}). We observe that increased $\epsilon$ is associated with a slight increase in objective risk comprehension (\linReg{0.03}{0}{0.06}{x}). As $\epsilon$ grows, the disparity between outcomes we provide (i.e., odds or sample DP outputs) also grows, which we hypothesize makes comparison easier. In addition, respondents who received the \ratiotext{} and \ratiovis{} explanations performed significantly better on these questions compared to respondents who received \samp{} (\linReg{0.57}{0.46}{0.68}{x}; \linReg{0.71}{0.60}{0.82}{x}). 

Our secondary analysis (Table~\ref{tab:secondary}) reveals that when adjusted for age, gender, and education (hereafter referred to as ``demographics''), higher numeracy skills are associated with higher objective comprehension (\linReg{0.29}{0.07}{0.52}{x}) and the highest level of baseline concern with increased objective comprehension (\linReg{0.27}{0.06}{0.48}{}). We posit this could be because highly concerned respondents may give more effort to understanding the presented information~\cite{celsi1988role}.

\begin{table*}[h]
\centering

\footnotesize

\begin{tabular}{lc|c||c|c} 
 \toprule
 \multicolumn{1}{c|}{\textbf{Variable}} & \multicolumn{1}{c|}{\textbf{\emph{Obj. Risk Comp.}}} 
 & \multicolumn{1}{c||}{\textbf{\emph{Obj. Risk Comp.}}} & \multicolumn{1}{c|}{\textbf{\emph{Subj. Privacy Und.}}} & \multicolumn{1}{c}{\textbf{\emph{Subj. Privacy Und.}}}\\
 
\multicolumn{1}{c|}{} & 
\multicolumn{1}{c|}{\scriptsize Expl. Ref: Deterministic} &
\multicolumn{1}{c||}{\scriptsize Expl. Ref: \samp{}} & 
\multicolumn{1}{c|}{\scriptsize Expl. Ref: Xiong et al.} &
\multicolumn{1}{c}{\scriptsize Expl. Ref: \samp{}}\\

\midrule
& $\beta$/CI & $\beta$/CI & OR/CI & OR/CI\\ 

 \midrule
 \multirow{2}{*}{Expl: \ratiotext} & $.14$ & $\mathbf{.57}^{***}$ & $1.61$ & $\mathbf{1.68}^{***}$ \\ 
& {\scriptsize [$-.04, .32$]} & {\scriptsize $[.46, .68]$} & {\scriptsize [$.96, 2.68$]} & {\scriptsize [$1.23, 2.28$]}\\

 \multirow{2}{*}{Expl: \ratiovis} & $\mathbf{.28}^{**}$ & $\mathbf{.71}^{***}$ & $1.46$ & $\mathbf{1.52}^{**}$ \\ 
& {\scriptsize [$.10, .46$]}& {\scriptsize [$.60, .82$]} & {\scriptsize [$.88, 2.43$]} & {\scriptsize [$1.11, 2.06$]}\\

 \multirow{2}{*}{Expl: \samp} & $\mathbf{-.43}^{***}$ & $-$ & $.97$ & $-$ \\ 
& {\scriptsize [$-.61, -.25$]}&  & {\scriptsize [$.58, 1.62$]} & \\

 \multirow{2}{*}{Setting: Optional} & $-.05$ & $-.04$ & $1.14$ & $1.19$ \\ 
& {\scriptsize [$-.13, .04$]}& {\scriptsize [$-.13, .06$]} & {\scriptsize [$.9, 1.45$]} & {\scriptsize [$.92, 1.52$]}\\

\multirow{2}{*}{$\epsilon$} & $-$ & $\mathbf{.03}^{*}$ & $-$ & $1.07$ \\ 
& & {\scriptsize [$0, .06$]} &  & {\scriptsize [$.98, 1.16$]}\\

\midrule
\end{tabular}
\caption{ \footnotesize \label{tab:objperceivedcomprehension} \emph{Left:} results from linear regression models examining relationships between number of correctly-answered objective-comprehension questions and experimental IVs. In the first column, the reference level for the explanation IV is the deterministic control, and in the second the reference is \samp{}. Note that dashes (i.e., $-$) in the first column indicate that $\epsilon$ was not included in the model. There are dashes in the second column for the \samp{} row because in this model, \samp{} was the reference level. We report regression coefficients $\beta$ and 95\% CIs for these coefficients. $\beta>0$ indicates an increase while $\beta<0$ indicates a decrease.
\emph{Right:} results from ordinal regression models examining relationships between subjective privacy understanding and IVs as on the left. Explanation reference levels are the Xiong et al. control (column 3) and \samp{} (column 4).
An OR $>1$ indicates an increase in odds, while an OR $<1$ indicates a decrease. * p$<0.05$; ** p$<0.01$; *** p$<0.001$.}
\end{table*}
\begin{table*}[!ht]
\centering
\footnotesize

\begin{tabular}{lc|c|c|c|c} 
 \toprule
 \multicolumn{1}{c|}{\textbf{Variable}} & \multicolumn{1}{c|}{\textbf{\emph{Obj. Risk Comp.}}} & \multicolumn{1}{c|}{\textbf{\emph{Subj. Privacy Und.}}} & \multicolumn{1}{c|}{\textbf{\emph{SE (Info.)}}} & \multicolumn{1}{c|}{\textbf{\emph{SE (Conf.)}}} & \multicolumn{1}{c}{\textbf{\emph{Willingness to Share}}} \\

\midrule
& $\beta$/CI & OR/CI & OR/CI & OR/CI & OR/CI \\

 \midrule
 
 \multirow{2}{*}{Age} 
 & $0$ 
 & $\mathbf{.99}^*$ 
 & $\mathbf{.99}^{**}$ 
 & $\mathbf{.99}^*$ 
 & $.99$ \\ 
 & {\scriptsize [$-.01, 0$]} 
 & {\scriptsize [$.98, 1.00$]}
 & {\scriptsize [$.98, .99$]} 
 & {\scriptsize [$.98, 1.00$]} 
 & {\scriptsize [$.98, 1.00$]} \\

  \multirow{2}{*}{Edu: Some College} 
  & $0$ 
  & $.85$ 
  & $1.05$ 
  & $1.17$ 
  & $.93$ \\ 
  & {\scriptsize [$-.16, .15$]} 
  & {\scriptsize [$.58, 1.25$]}
  & {\scriptsize [$.73, 1.51$]} 
  & {\scriptsize [$.81, 1.68$]} 
  & {\scriptsize [$.61, 1.43$]} \\

  \multirow{2}{*}{Edu: Bachelor's +} 
  & $.12$ 
  & $\mathbf{.66}^*$ 
  & $.80$ 
  & $.94$ 
  & $.71$ \\ 
  & {\scriptsize [$-.02, .26$]} 
  & {\scriptsize [$.46, .94$]}
  & {\scriptsize [$.57, 1.13$]} 
  & {\scriptsize [$.67, 1.33$]} 
  & {\scriptsize [$.47, 1.05$]} \\

  \multirow{2}{*}{Gender (man)} 
  & $-.02$ 
  & $\mathbf{1.65}^{***}$ 
  & $\mathbf{1.35}^{*}$ 
  & $\mathbf{1.32}^*$ 
  & $.91$ \\ 
  & {\scriptsize [$-.12, .08$]} 
  & {\scriptsize [$1.29, 2.11$]}
  & {\scriptsize [$1.07, 1.70$]} 
  & {\scriptsize [$1.05, 1.67$]} 
  & {\scriptsize [$.70, 1.19$]} \\

 \midrule
 
 \multirow{2}{*}{BC: Somewhat Concerned} 
 & $.15$ 
 & $.91$ 
 & $.75$ 
 & $\mathbf{.44}^{**}$ 
 & $\mathbf{.43}^*$ \\ 
 & {\scriptsize [$-.08, .38$]} 
 & {\scriptsize [$.51, 1.62$]}
 & {\scriptsize [$.42, 1.34$]} 
 & {\scriptsize [$.24, .80$]} 
 & {\scriptsize [$.18, .95$]} \\

 \multirow{2}{*}{BC: Concerned} 
 & $.18$ 
 & $.94$ 
 & $\mathbf{.56}^*$ 
 & $\mathbf{.31}^{***}$ 
 & $\mathbf{.31}^{**}$ \\ 
 & {\scriptsize [$-.04, .40$]} 
 & {\scriptsize [$.55, 1.62$]}
 & {\scriptsize [$.33, .98$]} 
 & {\scriptsize [$.17, .54$]} 
 & {\scriptsize [$.13, .66$]} \\

 \multirow{2}{*}{BC: Very Concerned} 
 & $\mathbf{.27}^*$ 
 & $1.27$ 
 & $.80$ 
 & $\mathbf{.39}^{**}$ 
 & $\mathbf{.19}^{***}$ \\ 
 & {\scriptsize [$.06, .48$]} 
 & {\scriptsize [$.75, 2.13$]}
 & {\scriptsize [$.47, 1.36$]} 
 & {\scriptsize [$.22, .69$]} 
 & {\scriptsize [$.08, .38$]} \\

  \multirow{2}{*}{Internet Skills} 
  & $-.02$ 
  & $\mathbf{1.39}^{***}$ 
  & $\mathbf{1.28}^{***}$ 
  & $1.15$ 
  & $.91$ \\ 
  & {\scriptsize [$-.08, .04$]} 
  & {\scriptsize [$1.19, 1.61$]}
  & {\scriptsize [$1.11, 1.47$]} 
  & {\scriptsize [$1.00, 1.32$]} 
  & {\scriptsize [$.77, 1.08$]} \\

  \multirow{2}{*}{Numeracy Skills} 
  & $\mathbf{.29}^*$ 
  & $1.06$ 
  & $.82$ 
  & $.79$ 
  & $.59$ \\ 
  & {\scriptsize [$.07, .52$]} 
  & {\scriptsize [$.6, 1.87$]}
  & {\scriptsize [$.48, 1.39$]} 
  & {\scriptsize [$.46, 1.36$]} 
  & {\scriptsize [$.31, 1.14$]} \\

  \midrule
\end{tabular}
\caption{ \label{tab:secondary} \footnotesize Results from our secondary analysis where we examined relationships between DVs and demographics (age, education, gender (man)) \& baseline concern (``BC''), internet skills, and numeracy skills. For interpretation of coefficients, see Table~\ref{tab:objperceivedcomprehension}.
}
\end{table*}

\subsubsection{Subjective Privacy Understanding (RQ1b)}
Next, we compare people's subjective privacy understanding of DP guarantees when presented with our explanations versus the Xiong et al. control (Table \ref{tab:objperceivedcomprehension}, right), but do not find that any of our explanation methods have significant effects versus the control. We construct a second model where \samp{} is the explanation method reference level. Here we find that our \ratiotext{} and \ratiovis{} explanation methods are associated with increased perceptions of understanding the privacy protection compared to \samp{} (\ordRegFirst{1.68}{1.23}{2.28}{}; \ordReg{1.52}{1.11}{2.06}{}).

In our secondary analysis (Table~\ref{tab:secondary}), we find that the highest level of education is associated with lower subjective privacy understanding (\ordReg{0.66}{0.46}{0.94}{x}), while identifying as a man is associated with higher subjective understanding (\ordReg{1.65}{1.29}{2.11}{x}). There is also a small effect of increased age on lower subjective understanding (\ordReg{0.99}{0.98}{1.00}{x}). Finally, when adjusting for demographics, we find that increased internet skills are associated with an increase in subjective understanding (\ordReg{1.39}{1.19}{1.61}{x}). Consistent with prior work, these findings suggest multiple social factors mediate people's perceptions of their understanding of privacy guarantees. For example, older adults may have lower confidence in their knowledge about S\&P topics~\cite{madden2017privacy}, men may report higher self-confidence across a variety of domains including digital skills and use~\cite{lundeberg1994highly,ross2012gender,pajares2002gender,chu2010family,hargittai2006differences}, and those with higher internet skills may perceive themselves as having greater understanding of digital concepts~\cite{hargittai2006differences}.\looseness=-1

\subsubsection{Self-Efficacy (RQ1c)}\label{sec:results:selfefficacy}
We measure self-efficacy in terms of (1) the extent to which respondents feel they have \textit{enough information to make these decisions} and (2) the extent to which they feel \textit{confident in making data-sharing decisions} (Table~\ref{tab:selfefficacy}). 

\smallskip
\noindent\textbf{Enough Information to Decide.}
We find that \samp{} have a significant negative effect on feelings of having enough information to decide when compared to the deterministic control (\ordReg{0.48}{0.30}{0.78}{x}). We hypothesize that respondents felt that the information presented in \samp{} misaligned with key pieces of information they needed. For example, one respondent wrote that they ``\textit{would like to know the likelihood that the [figure in the] report [the manager] receives is higher or lower}'' (than the total number of \no{} responses), indicating they may have wanted a summary of probabilities like in the \ratiotext{} or \ratiovis{} explanations. Another respondent wrote: ``\textit{I would want to know the chances that my [manager] gets a higher number.}'' 

Compared to the Xiong et al. control group, respondents who received the \ratiotext{} explanation were over 75\% more likely to report feeling a point higher in our 4-point semantic scale for having enough information to decide (\ordReg{1.76}{1.07}{2.88}{x}) and respondents who received the \ratiovis{} explanation were over 65\% more likely (\ordReg{1.66}{1.02}{2.72}{x}). On the other hand, no such association was found for \samp{}.

Compared to the \samp{} respondents, we find that participants who received the \ratiotext{} and \ratiovis{} explanation methods were over 65\% and 55\% more likely, respectively, to report feeling a point higher on the scale for having enough information to decide (\ordReg{1.66}{1.22}{2.26}{}; \ordReg{1.57}{1.16}{2.12}{}). We do not find a significant effect of $\epsilon$ on feelings of having enough information to decide. Finally, our secondary analysis (Table~\ref{tab:secondary}) indicates a small effect of increased age on decreased feelings of enough information to decide (\ordReg{0.99}{0.98}{0.99}{x}) and that identifying as a man is associated with higher such feelings (\ordReg{1.35}{1.07}{1.70}{x}). When adjusting for demographics, increased internet skills are also associated with higher such feelings (\ordReg{1.28}{1.11}{1.47}{}) and baseline concern of ``concerned'' is associated with lower such feelings (\ordReg{0.56}{0.33}{0.98}{}).

\smallskip
\noindent\textbf{Confidence Deciding.}
We do not find significant relationships between our experimental explanations and confidence deciding when compared to either the deterministic control or the Xiong et al. control. For the third model where we hold \samp{} as the explanation reference, we do not find that $\epsilon$ has a significant effect on confidence deciding. However, our \ratiotext{} and \ratiovis{} explanations are associated with an increase in feelings of confidence deciding compared to the \samp{} method (\ordReg{1.58}{1.16}{2.15}{x}; \ordReg{1.50}{1.11}{2.04}{x}). Through our secondary analysis (Table~\ref{tab:secondary}), we find that being a man is associated with higher feelings of confidence deciding (\ordReg{1.32}{1.05}{1.67}{x}). We also find and that all three higher levels of baseline concern (adjusted for demographics) are associated with lower confidence deciding. We posit that those more concerned about negative repercussions may feel more conflicted about which data-sharing decision to make.

\begin{table*}[!ht]
\centering
\footnotesize

\begin{tabular}{lc|c|c||c|c|c} 
 \toprule
 \multicolumn{1}{c|}{\textbf{Variable}} & \multicolumn{1}{c|}{\textbf{\emph{SE (Info.)}}} & \multicolumn{1}{c|}{\textbf{\emph{SE (Info.)}}} & \multicolumn{1}{c||}{\textbf{\emph{SE (Info.)}}} & \multicolumn{1}{c|}{\textbf{\emph{SE (Conf.)}}} & \multicolumn{1}{c|}{\textbf{\emph{SE (Conf.)}}} & \multicolumn{1}{c}{\textbf{\emph{SE (Conf.)}}} \\
 
\multicolumn{1}{c|}{} & \multicolumn{1}{c|}{\scriptsize Expl. Ref: Det.} & \multicolumn{1}{c|}{\scriptsize Expl. Ref: Xiong et al.} & \multicolumn{1}{c||}{\scriptsize Expl. Ref: SR} & \multicolumn{1}{c|}{\scriptsize Expl. Ref: Det.}& \multicolumn{1}{c|}{\scriptsize Expl. Ref: Xiong et al.} & \multicolumn{1}{c}{\scriptsize Expl. Ref: SR}\\ 

\midrule
& OR/CI & OR/CI & OR/CI & OR/CI & OR/CI  & OR/CI\\

 \midrule
 \multirow{2}{*}{Expl: \ratiotext} 
    & $.80$ 
    & $\mathbf{1.76}^*$ 
    & $\mathbf{1.66}^{**}$ 
    & $1.55$ 
    & $1.19$ 
    & $\mathbf{1.58}^{**}$ \\ 
    & {\scriptsize [$.49, 1.3$]} 
    & {\scriptsize [$1.07, 2.88$]}
    & {\scriptsize [$1.22, 2.26$]} 
    & {\scriptsize [$.94, 2.55$]} 
    & {\scriptsize [$.73, 1.93$]} 
    & {\scriptsize [$1.16, 2.15$]}\\

 \multirow{2}{*}{Expl: \ratiovis} 
 & $.76$ 
 & $\mathbf{1.66}^*$ 
 & $\mathbf{1.57}^{**}$ 
 & $1.47$ 
 & $1.13$
 & $\mathbf{1.50}^{**}$  \\ 
 & {\scriptsize [$.47, 1.22$]} 
 & {\scriptsize [$1.02, 2.72$]}
 & {\scriptsize [$1.16, 2.12$]} 
 & {\scriptsize [$.89, 2.42$]} 
 & {\scriptsize [$.70, 1.83$]} 
 & {\scriptsize [$1.11, 2.04$]} \\

 \multirow{2}{*}{Expl: \samp} 
 & $\mathbf{.48}^{**}$ 
 & $1.07$ 
 & $-$ 
 & $.99$  
 & $.75$ 
 & $-$  \\ 
 & {\scriptsize [$.30, .78$]} 
 & {\scriptsize [$.65, 1.74$]} 
 &  
 & {\scriptsize [$.60, 1.62$]} 
 & {\scriptsize [$.47, 1.22$]} 
 &  \\

 \multirow{2}{*}{Setting: Optional} 
 & $.98$ 
 & $.99$ 
 & $1.00$ 
 & $1.10$ 
 & $1.09$ 
 & $1.10$ \\ 
 & {\scriptsize [$.77, 1.25$]} 
 & {\scriptsize [$.78, 1.26$]} 
 & {\scriptsize [$.78, 1.28$]} 
 & {\scriptsize [$.87, 1.4$]} 
 & {\scriptsize [$.86, 1.38$]} 
 & {\scriptsize [$.86, 1.41$]} \\

 \multirow{2}{*}{$\epsilon$} 
 & $-$ 
 & $-$ 
 & $1.08$
 & $-$ 
 & $-$ 
 & $1.07$ \\ 
 & 
 &  
 & {\scriptsize [$1.00, 1.18$]} 
 &  
 &  
 & {\scriptsize [$.99,1.16$]} \\

  \midrule
\end{tabular}
\caption{ \label{tab:selfefficacy} \footnotesize Ordinal regression models examining ``enough info'' (Info.) and ``confidence deciding'' (Conf).
SR=\samp{}; Det.=Deterministic control. See Table~\ref{tab:objperceivedcomprehension} for interpretation.
}
\end{table*}

\subsection{Influence on Data Sharing (RQ2)}

To answer RQ2, we investigate the extent to which our experimental explanations influence people's data-sharing decisions (Table~\ref{tab:willingnesstoshare}).
Compared to both the deterministic and Xiong et al. controls, the \samp{} and \ratiotext{} explanations have significant relationships with willingness to share data. Respondents are about 2 times as likely to share their data when shown the \ratiotext{} explanation compared to the deterministic control (\ordRegHalf{1.14}{3.42}{}), and about 3 times as likely to share their data when shown \samp{} compared to the deterministic control (\ordRegHalf{1.73}{5.22}{}), an interesting finding considering the \samp{} explanation did not seem to well-support respondents in objective risk comprehension (see Section \ref{sec:results:objcomp}). Respondents were over twice as likely to share their data if shown the \ratiotext{} explanation (\ordRegHalf{1.41}{4.3}{}) compared to the Xiong et al. control, nearly two times as likely when shown the \ratiovis{} explanation (\ordRegHalf{1.12}{3.37}{}), and nearly four times as likely when shown the \samp{} explanation (\ordRegHalf{2.14}{6.57}{}).

\begin{floatingfigure}{9cm}
    \vspace*{-3mm}
    \centering
    \includegraphics[width=.55\textwidth]{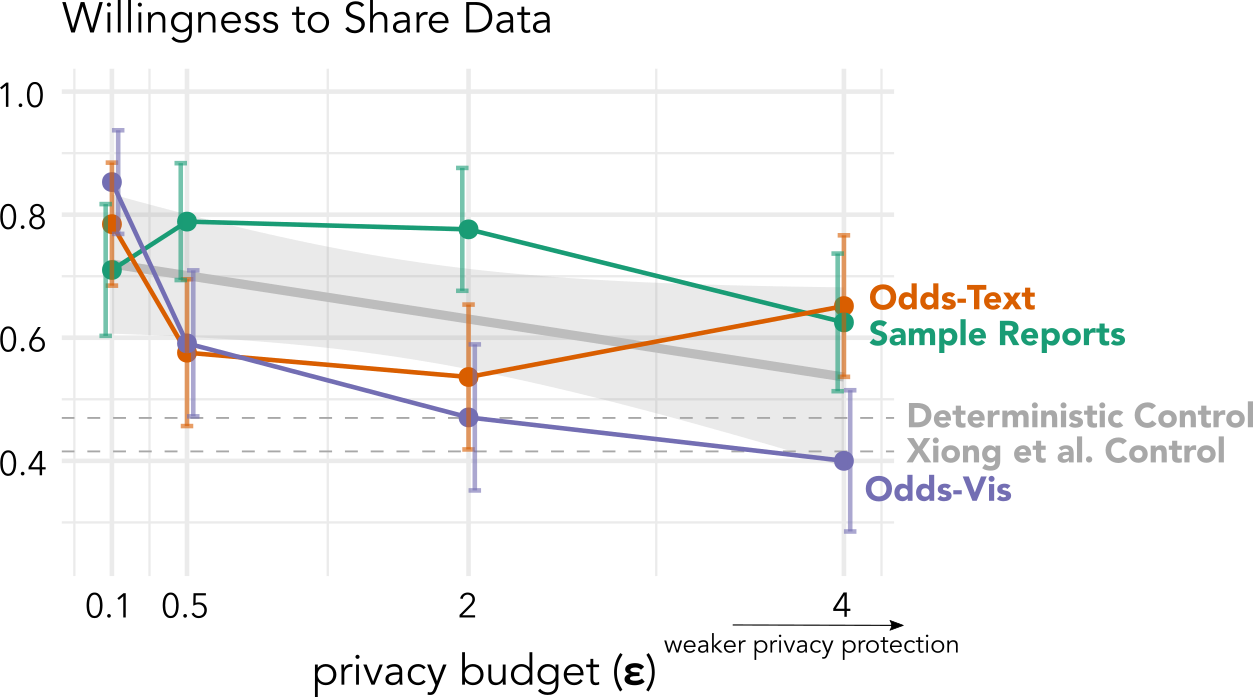}
    \vspace*{-5mm}
    \captionsetup{labelformat=empty}
    \caption{Figure~\ref{fig:willing-to-share}.\footnotesize Proportion of respondents willing to share data across explanation methods and $\epsilon$, shown with 95\% binomial CIs. We plot a regression line (solid gray) between proportion of data sharing across our methods and $\epsilon$.}
    \vspace*{-2mm}
    \label{fig:willing-to-share}
\end{floatingfigure}

We additionally construct a logistic regression model to examine the relationship between the DV, willingness to share data, and $\epsilon$. We find that as $\epsilon$ increases, and privacy protections become weaker, respondents are less likely to share their data (\ordReg{0.81}{0.74}{0.89}{}). Figure~\ref{fig:willing-to-share} shows the proportion of respondents who said that they would share their data, for each explanation method and $\epsilon$ value. Furthermore, we find that respondents are less likely to share data if given the \ratiotext{} or \ratiovis{} explanations than the \samp{} explanation (\ordReg{0.65}{0.45}{0.94}{} and \ordReg{0.51}{0.35}{0.73}{}). We hypothesize that this may be related to differences in feelings of self-efficacy between the explanations (see Section~\ref{sec:results:selfefficacy}). In the model with \samp{} as reference, we also find that compared to answering truthfully in the mandatory setting, people are less likely to share their data in the optional setting (\ordReg{0.74}{0.55}{0.99}{})---in the mandatory setting, people may feel uncomfortable lying. When asked to explain their decision-making, many participants explicitly expressed a desire to be honest. For example, one respondent wrote that they ``\textit{would tell the truth regardless because [they] refuse to lie regardless of the outcome.}'' Finally, when adjusting for demographics, the higher three baseline concern levels understandably have a small decreased effect on sharing data (Table \ref{tab:secondary}).

The open-text responses provide further context on respondents' decision-making. Many respondents explicitly reasoned about how their behavior would change the odds of their manager believing they responded \no. For example, one respondent wrote: ``\textit{The chance that the manager will believe I responded no is only slightly higher if I participate than if I don't, so I may as well give my opinion.}'' Others expressed less concern about privacy and instead focused on the utility of the collected data. For example, one respondent argued that participating in the survey was ``\textit{the right thing to do,}'' since it might lead to improved conditions for their coworkers.

\begin{table*}[t]
\centering

\footnotesize

\begin{tabular}{lc|c|c} 
 \toprule
 \multicolumn{1}{c|}{\textbf{Variable}} & \multicolumn{1}{c|}{\textbf{\emph{Willingness to Share}}} & \multicolumn{1}{c|}{\textbf{\emph{Willingness to Share}}} & \multicolumn{1}{c}{\textbf{\emph{Willingness to Share}}} \\
 
\multicolumn{1}{c|}{} & 
\multicolumn{1}{c|}{\scriptsize Expl. Ref: Deterministic} &
\multicolumn{1}{c|}{\scriptsize  Expl. Ref: Xiong et al.} &
\multicolumn{1}{c}{\scriptsize Expl. Ref: \samp{}}\\

\midrule
& OR/CI  & OR/CI  & OR/CI \\ 

 \midrule
 \multirow{1}{*}{Expl: \ratiotext{}} 
 & $\mathbf{1.97}^*$ 
 & $\mathbf{2.45}^{**}$ 
 & $\mathbf{.65}^*$  \\ 
 & {\scriptsize [$1.14, 3.42$]} 
 & {\scriptsize [$1.41,4.3$]} 
 & {\scriptsize [$.45,.94$]}\\

 \multirow{1}{*}{Expl: \ratiovis{}} 
 & $1.55$ 
 & $\mathbf{1.93}^*$ 
 & $\mathbf{.51}^{***}$ \\ 
 & {\scriptsize [$.9, 2.68$]} 
 & {\scriptsize [$1.12,3.37$]} 
 & {\scriptsize [$.35,.73$]}\\

 \multirow{1}{*}{Expl: \samp{}} 
 & $\mathbf{2.99}^{***}$ 
 & $\mathbf{3.72}^{***}$ 
 & $-$  \\ 
 & {\scriptsize [$1.73, 5.22$]} 
 & {\scriptsize [$2.14,6.57$]} 
 & \\

 \multirow{1}{*}{Setting: Optional} 
 & $.80$ 
 & $.76$ 
 & $\mathbf{.74}^*$  \\ 
 & {\scriptsize [$.61, 1.06$]} 
 & {\scriptsize [$.57, 1$]} 
 & {\scriptsize [$.55, .99$]}\\

 \multirow{1}{*}{$\epsilon$} 
 & $-$ 
 & $-$ 
 & $\mathbf{.81}^{***}$ \\ 
 &  
 &  
 & {\scriptsize [$.74,.89$]}\\

  \midrule
\end{tabular}

 \caption{\footnotesize Logistic regression models examining relationships between willingness to share data and our IVs. For table interpretation, see Table~\ref{tab:objperceivedcomprehension}.}
 
\label{tab:willingnesstoshare}
\end{table*}

\section{Discussion}
\label{sec:discussion}

We find that the odds-based explanation methods (\ratiovis{} in all cases and \ratiotext{} in most) improve objective risk comprehension, subjective privacy understanding, and self-efficacy compared to the examples-based explanation method (\samp{}). The odds-based methods also improve feelings of having enough information to make privacy decisions over an existing state-of-the-art DP description~\cite{SP:XWLJ20}, which does not offer information about $\epsilon$, suggesting that providing more detail about privacy protection---even probabilistic information with which people may struggle~\cite{slovic2000perception}---improves some forms of self-efficacy in data-sharing decisions. Further, those offered information about $\epsilon$ are also more willing to share data than who are offered privacy protection described without information about $\epsilon$. However, those given \samp{} were more likely to share data compared to those given either odds-based explanation, despite their comparatively poor objective risk comprehension.

Finally, while it is encouraging that respondents' data-sharing decisions change when presented with privacy-protecting information, we also note that their data-sharing decisions, objective understanding of the risk of those decisions, and their confidence in making those decisions all depend on how concerned they are about the consequence of their decision. It may be that people who are particularly concerned invest more cognitive resources to accurately reason about probabilistic privacy guarantees.\footnote{See \cite{acquistinudges} for cognitive biases that may influence data-sharing decisions.}

\smallskip
\noindent
    \textbf{Communicating Utility Implications.}
Our work builds on prior work illustrating that odds can effectively communicate privacy risk in comparison with a no-privacy or no-data-collection alternative~\cite{kaptchuk2020good,franzen2022private} to show that explanation methods that communicate odds can help people better understand \textit{probabilistic} privacy guarantees. 
However, our qualitative results indicate that, in at least some contexts, people may also be concerned about utility implications of DP. For example, some respondents felt that it was important for their input to be faithfully communicated, e.g., to improve their workplace environment. Higher amounts of DP noise, while affording stronger privacy protections, reduce the accuracy and ``utility'' of released statistics.
At face value it may seem that accuracy concerns are more in the domain of data curators (e.g., the interfaces described in Section~\ref{sec:relatedwork:decisionmaking}), but we suggest that people may similarly require depictions of accuracy to make effective judgments about the downstream utility of their data. This in turn supports people to make informed decisions about data sharing. In line with prior research illustrating that presenting information about both accuracy and privacy improves ability to predict people's data-sharing decisions in medical contexts~\cite{kaptchuk2020good} and prior work aiming to convey accuracy implications of DP to people~\cite{xiong2022using}, people may benefit from reasoning not only about privacy, but also about the accuracy-privacy trade-off in the DP context. Thus, we emphasize the need to go ``beyond'' privacy to consider utility as a key part of respectful data use~\cite{redmiles2021need}. \looseness=-1 
Our results yield insight into \textit{how} utility implications may be most effectively communicated to people.
Our \ratiotext{} and \ratiovis{} explanation methods demonstrate how to map $\epsilon$ to outcome probabilities. It may be additionally useful for explanations to provide mappings from $\epsilon$ to utility-related outcomes. 
For example, if receiving a certain number of NO responses will require the manager to complete additional training, odds-based methods can similarly be used to communicate the probability that that threshold value will be met given a person's response under a given $\epsilon$. Such information could help them assess utility of their response in terms of leading to a tangible outcome.

\smallskip
\noindent
    \textbf{Toward More Complex Scenarios.}
Our work contributes a framework for communicating $\epsilon$ implications to end users; details about our specific scenario can be changed to apply our methods to varied applications of DP. For example, we do not explicitly model the adversary's prior belief about an individual responding in a negative way, however our framework can be easily applied to a scenario where the prior is specified and the adversary performs a Bayesian update. We provide details in Appendix~\ref{appendix:priors}. To apply our odds-based explanation methods to other types of queries, such as mean queries, we suggest computing probabilities based on a hypothesis testing framework~\cite{liu2019investigating}. Finally, as future work, we see promise in creating guidance and tooling that allows people employing DP to generate explanations consistent with specifics of their application settings. Such work would encourage practical use of explanation methods that clarify the impact of $\epsilon$.

\smallskip
\noindent
    \textbf{Public Deliberation Around $\epsilon$.}
Dwork et al.~\cite{dwork2019differential} have proposed a registry of $\epsilon$ values (and other implementation details) used by organizations applying DP. They argue that such a registry could enable comparisons across differentially-private systems and increase accountability around the use of DP, lowering chances of privacy theater. Our explanations can increase the impact of such a registry by building on work that aims to translate the implications of $\epsilon$ to a wider audience, thus helping facilitate public deliberation around privacy budgets. More generally, our explanation methods and other methods of translating $\epsilon$ can support external audits by making implementation decisions like $\epsilon$ more widely interpretable and easier to discuss among interested parties with a diverse set of expertise and backgrounds. For example, recent debates around the U.S. Census Bureau's use of DP for the 2020 Census demonstrated challenges of effectively discussing aspects of DP among several groups (computer scientists, policymakers, demographers, non-profit organizations, the public, etc.)~\cite{nanayakkara2022s,boyd2022differential}. These challenges arose in part because the technical specifics of DP (such as privacy budgets) can be abstract or counterintuitive, especially when discussed outside the computer science literature. Explanations of $\epsilon$ effective for a broad population can close gaps in communication and encourage public deliberation over privacy policies.

\section{Acknowledgements}
We would like to thank Elizabeth Chase and Sean Kross for guidance on our analysis, Jessica Hullman for reviewing an earlier draft of the paper, Frauke Kreuter for feedback on survey design, and Chase Stokes for visualization feedback. In addition, we thank the attendees of Theory and Practice of Differential Privacy (TPDP) 2022, attendees of the Symposium on Applications of Contextual Integrity 2022, members of the MU Collective at Northwestern University, and members of a Columbia University privacy reading group for valuable discussion and feedback.

All authors were supported by DARPA (contract number W911NF-21-1-0371). Any opinions, findings, and conclusions or recommendations expressed in this material are those of the authors and do not necessarily reflect the views of the United States Government or DARPA. In addition to DARPA support, the third author was supported in part by NSF grant CNS-1942772 (CAREER), a Mozilla Research Grant, a JPMorgan Chase Faculty Research Award, and an Apple Privacy-Preserving Machine Learning Award. The fourth author was also supported by NSF grant \#2030859 to the Computing Research Association for the CIFellows Project, and the fifth author was also supported by a Google Research Scholar Award.

\bibliographystyle{plain}
\bibliography{references,abbrev3,crypto}  

\appendix
\newpage
\section{Demographics}
Table~\ref{tab:demographics} provides a breakdown of survey respondent demographics. Since some survey respondents declined to answer demographic questions and some selected multiple answers for race/ethnicity/gender, the counts for each category do not always add up to 963. Also note that while only five respondents explicitly wrote-in ``biracial,'' ``multi-racial,'' or ``mixed,'' many respondents selected more than one answer for the race/ethnicity question. Respondent age was calculated by subtracting the provided birth year from the year the survey was taken.
\label{app:demographics}

\begin{table}[htbp]
\caption{Survey Respondent Demographics}
\begin{center}
\small
\begin{tabular}{l r}
\hline
\textbf{Demographic Attribute} & \textbf{Count} \\
\hline
\multicolumn{2}{l}{\textbf{\emph{Gender}}} \\ 
\hline
Fluid gender & 1 \\
Man & 479 \\
Non-binary & 24 \\
Woman & 458 \\
Prefer not to answer & 5\\
\hline
\multicolumn{2}{l}{\textbf{\emph{Age}}} \\ 
\hline 
18-29 & 346 \\
30-39 & 286 \\
40-49 & 142 \\
50+ & 173 \\
Prefer not to answer & 16\\
\hline
\multicolumn{2}{l}{\textbf{\emph{Race/Ethnicity}}} \\ 
\hline
Hispanic or Latino & 104\\
Black or African American & 114 \\
White & 718 \\
American Indian or Alaska Native & 19\\
Asian, Native Hawaiian, or Pacific Islander & 84\\
Mixed, Multiracial, or Biracial & 5\\
Unique free-text responses & 3 \\
Prefer not to answer & 9 \\
\hline
\multicolumn{2}{l}{\textbf{\emph{Education}}} \\ 
\hline
High school or less & 140 \\
Some college & 319 \\
Bachelor’s or above & 501\\
Prefer not to answer & 3\\
\hline
\multicolumn{2}{l}{\textbf{\emph{Education/work in computer science/IT}}} \\
\hline
Yes & 192\\
No & 735 \\
Prefer not to answer & 36 \\
\hline
\multicolumn{2}{l}{\textbf{\emph{Income}}} \\
\hline
Less than \$10,000 & 60\\
\$10,000 to under \$20,000 & 64\\
\$20,000 to under \$30,000 & 89\\
\$30,000 to under \$40,000 & 95\\
\$40,000 to under \$50,000 & 79 \\
\$50,000 to under \$65,000 & 131\\
\$65,000 to under \$80,000 & 103\\
\$80,000 to under \$100,000 & 93\\
\$100,000 to under \$125,000 & 77\\
\$125,000 to under \$150,000 & 49\\
\$150,000 to under \$200,000 & 38\\
\$200,000 or more & 42\\
Prefer not to answer & 43 \\
\hline
\end{tabular}
\label{tab:demographics}
\end{center}
\end{table}

\newpage

\section{Codebook}
\label{app:codebook}

Table~\ref{tab:codebook} shows the codebook used for qualitative analysis.

\begin{table}[htbp]
\caption{Codebook}
\begin{center}
\small
\begin{tabular}{p{0.36\linewidth}  p{0.53\linewidth}}
\hline
\textbf{Code} & \textbf{Example} \\
\hline
Confusion/questions about privacy protection & \textit{It's a bit confusing to understand but I think it would send out various reports on the number of people who said no} \\
\hline
Utility of response & \textit{The random process completely obfuscates the true number of NO responses; that is great for employee anonymity, but is kind of useless for the manager.} \\
\hline
Odds comparison & \textit{There are greater odds that if I respond ``no", my manager will get a report that leads them to believe I responded ``no".} \\
\hline
Scenario context & \textit{Who else gets the survey results? Does HR get the correct information and so I trust that HR will help me if the boss retaliates}  \\
\hline
Honesty & \textit{I've actually been in a somewhat similar situation, and I was retaliated against. Still, despite that experience, I think it is important to me to be honest about what is going on.} \\
\hline
\end{tabular}
\label{tab:codebook}
\end{center}
\end{table}

\section{Survey Instrument}
\label{appendix:surveyinstrument}

\subsection{Scenario Text}
\label{appendix:surveyinstrument:scenariotext}

The following is the scenario text for the mandatory setting:

\vspace{.25cm}
\hrule
\vspace{.25cm}

Imagine you work on a team with four other people. All five of you report to the same manager. The company is \underline{requiring} each of you to participate in a survey. The survey asks the following \yes/\no{} question:

Do you feel adequately supported by your manager?

You have had negative experiences with your manager and want to answer \no.

However, you don’t want your manager to find out you responded \no. Your manager may retaliate if they believe you responded \no. For example, they might give you a negative performance review, assign you extra work, or try to get you fired.

You must decide how to respond to the survey question. You can respond truthfully by responding \no or respond untruthfully by responding \yes.

\vspace{.25cm}
\hrule
\vspace{.25cm}

Based on lunchtime conversations, it is obvious to your manager that all your other teammates will respond \yes, indicating they feel adequately supported by your manager. On the other hand, your manager has no idea how you will respond.

Your manager will receive a report on the results of the survey. This report will say the total number of people who responded \no.

Even though the report will keep your names anonymous, once your manager gets the report, they can still use it to guess your response.

For example, imagine the report shows that there was one \no{} response. Your manager will believe it was you because they believe your other teammates all responded \yes.

\textit{(The deterministic control ends here.)}

\textit{(If the Xiong et al. control is shown, it appears here, and the scenario ends afterward.)}

\vspace{.25cm}
\hrule
\vspace{.25cm}

However, your company will use a privacy protection method to help prevent your manager from correctly guessing anyone’s response.

Your company will not report \underline{exactly} how many employees on your team responded \no. Instead, they will generate many potential reports by using a statistical method to modify the total number of \no{} responses. So, each potential report may show a number somewhat lower or higher than the actual number of \no{} responses.

\textit{Only \underline{ONE} report will be randomly sent to your manager.}

Note that due to the privacy method, it is possible that the specific report your manager receives will lead them to believe that you responded \no{} regardless of what you respond on the survey.

\vspace{.25cm}
\hrule
\vspace{.25cm}

\textit{(\ratiotext{}, \ratiovis{}, or \samp{} appears here.)}\\

\vspace{.25cm}
\hrule
\vspace{.25cm}

\noindent
\textbf{Willingness to Share:}
\begin{itemize}
    \item Based on the scenario and description of privacy protection, would you respond \no{} (i.e., respond truthfully) to the survey question? (Yes/No/I prefer not to answer this question.)
    \item Briefly explain your reasoning. (open-text response)
\end{itemize}

\noindent
\textbf{Objective Risk Comprehension:}
\begin{itemize}
    \item My manager is more likely to believe I responded \no{} if I respond \no{} (i.e., respond truthfully) to the survey question than if I respond \yes{} (i.e., respond untruthfully) to the survey question. (True/False/I don't know/I prefer not to answer this question.)
    \item My manager is more than twice as likely to believe I responded \no{} if I respond \no{} (i.e., respond truthfully) to the survey question than if I respond \yes{} (i.e., respond untruthfully) to the survey question. (True/False/I don't know/I prefer not to answer this question.)
\end{itemize}

\noindent
\textbf{Subjective Privacy Understanding:}
\begin{itemize}
    \item Based on the scenario and description of the privacy protection, how confident are you that you understand the privacy protection applied to the survey results? (Not at all confident/Somewhat confident/Confident/Very confident/I prefer not to answer this question.)
\end{itemize}

\noindent
\textbf{Self-Efficacy:}
\begin{itemize}
    \item Based on the scenario and description of the privacy protection, how confident are you that you have \underline{enough information to decide} whether to respond \no{} (i.e., respond truthfully) to the survey question? (Not at all confident/Somewhat confident/Confident/Very confident/I prefer not to answer this question.)
    \item What further information, if any, would you like to have to help you decide whether you would respond \no{} (i.e., respond truthfully) to the survey question? (open-text response)
    \item Based on the scenario and the description of the privacy protection, how confident are you in \underline{deciding} whether to respond \no{} (i.e., respond truthfully) to the survey question? (Not at all confident/Somewhat confident/Confident/Very confident/I prefer not to answer this question.)
\end{itemize}

\subsection{Xiong et al.~\cite{SP:XWLJ20} Control Text}
\label{sec:surveyinstrument:xiongbaseline}
\begin{quote}
    \textit{However, to respect your personal information privacy, the report shared with your manager will include the total number of \no{} responses processed using a privacy protection method. This method protects employees’ privacy by adding random noise to aggregated data, for example, the total number of \no{} responses, such that the probability that your manager can infer your response on the survey is lower than without the privacy protection.}
\end{quote}

\newpage
\subsection{\samp{} Values}
\label{appendix:surveyinstrument:samplereports}

Values used for \samp{} under each of our four selected privacy budgets:

\begin{table}[h]
\small
\begin{tabular}{|l||lllll|}
\hline
& \multicolumn{5}{l|}{\textbf{If you do not participate/respond \yes}}                                                    
\\ \hline
$\epsilon$ & \multicolumn{1}{l|}{\#1}   & \multicolumn{1}{l|}{\#2}     & \multicolumn{1}{l|}{\#3}   & \multicolumn{1}{l|}{\#4}    & \#5  
\\ \hline
$0.1$     & \multicolumn{1}{l|}{$4.0$} & \multicolumn{1}{l|}{-$12.2$} & \multicolumn{1}{l|}{$2.4$} & \multicolumn{1}{l|}{$23.7$} & -$13.5$
\\ \hline
$0.5$     & \multicolumn{1}{l|}{$0.8$} & \multicolumn{1}{l|}{-$2.4$}  & \multicolumn{1}{l|}{$0.5$} & \multicolumn{1}{l|}{$4.7$}  & -$2.7$ 
\\ \hline
$2$       & \multicolumn{1}{l|}{$0.2$} & \multicolumn{1}{l|}{-$0.6$}  & \multicolumn{1}{l|}{$0.1$} & \multicolumn{1}{l|}{$1.2$}  & -$0.7$ 
\\ \hline
$4$       & \multicolumn{1}{l|}{$0.1$} & \multicolumn{1}{l|}{-$0.3$}  & \multicolumn{1}{l|}{$0.1$} & \multicolumn{1}{l|}{$0.6$}  & -$0.3$ 
\\ \hline
\hline
& \multicolumn{5}{l|}{\textbf{If you participate/respond \no}}                              
\\ \hline
$\epsilon$ & \multicolumn{1}{l|}{\#1}   & \multicolumn{1}{l|}{\#2}   & \multicolumn{1}{l|}{\#3}     & \multicolumn{1}{l|}{\#4}     & \#5   
\\ \hline
$0.1$     & \multicolumn{1}{l|}{$7.1$} & \multicolumn{1}{l|}{$3.4$} & \multicolumn{1}{l|}{-$14.8$} & \multicolumn{1}{l|}{-$23.0$} & $0.0$ 
\\ \hline
$0.5$     & \multicolumn{1}{l|}{$2.2$} & \multicolumn{1}{l|}{$1.5$} & \multicolumn{1}{l|}{-$2.2$}  & \multicolumn{1}{l|}{-$3.8$}  & $0.8$ 
\\ \hline
$2$       & \multicolumn{1}{l|}{$1.3$} & \multicolumn{1}{l|}{$1.1$} & \multicolumn{1}{l|}{$0.2$}   & \multicolumn{1}{l|}{-$0.2$}  & $1.0$ 
\\ \hline
$4$       & \multicolumn{1}{l|}{$1.2$} & \multicolumn{1}{l|}{$1.1$} & \multicolumn{1}{l|}{$0.6$}   & \multicolumn{1}{l|}{$0.4$}   & $1.0$ 
\\ \hline
\end{tabular}
\end{table}

\section{DAG}
\label{appendix:DAG}
\begin{figure}[h]
  \includegraphics[width=.7\linewidth]{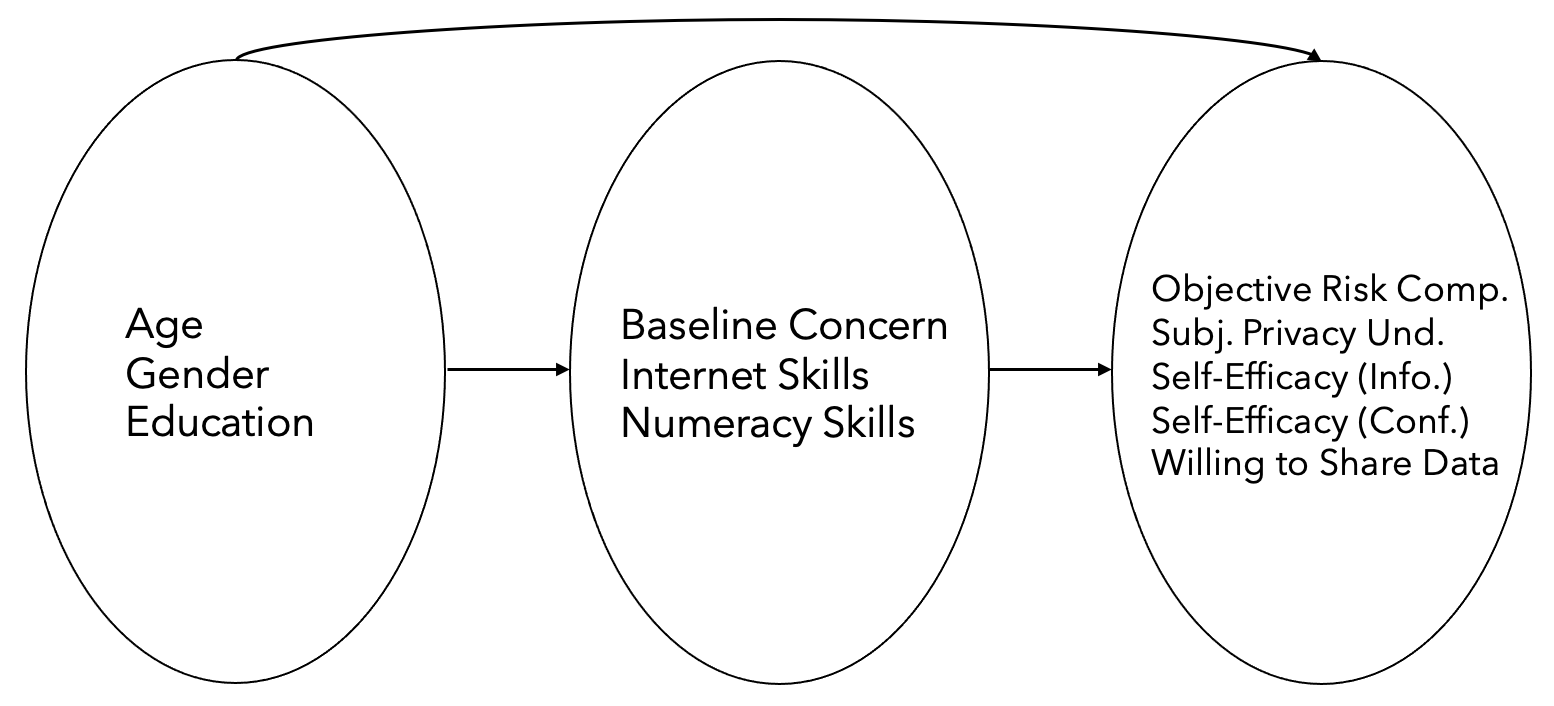}
    \caption{\footnotesize The diagram depicts the directed acyclic graphs (DAGs) we used to inform our analysis.}
  \label{fig:DAG}
\end{figure}

\section{Modeling Adversary's Priors}
\label{appendix:priors}
To make the scenario discussed in this paper more realistic and flexible, we can account for the fact that the manager may have a prior belief about how the respondent will respond even before seeing the privacy-protected report. We can model the manager's thought process by specifying their prior belief and supposing that they perform a Bayesian update to obtain a posterior belief based on the DP output in the report. We show that under this model, changing the manager's prior is equivalent to shifting the threshold ($r_{\text{threshold}}$) on the DP output at which the manager believes that one outcome (i.e., the respondent said \no{}) is more likely than the other (i.e., the respondent did not say \no{}). 

If the respondent answers \no{}, the manager will see a sample drawn from a Laplace distribution centered at $1$. Otherwise, the manager sees a sample drawn from a Laplace distribution centered at $0$. Thus, the manager's task is to guess whether the DP output $r$ was drawn from $Lap(\mu=1, b =  \frac{1}{\epsilon})$ or $Lap(\mu=0, b =  \frac{1}{\epsilon})]$.

We use Bayes’ Theorem to calculate the probability that the manager finds it more likely that the respondent answered \no{} after viewing a DP output $r$ (i.e., the manager's posterior). Let $f_i$ denote the probability density function for $Lap(\mu=i, b =  \frac{1}{\epsilon})]$, and let $P_{no}$ denote the manager's prior belief that the respondent answered \no{}. The updated probability that the respondent answered \no{} (the posterior probability) is given by:
\begin{equation*}
    \frac{f_1(r) P_{no}}{f_1(r) P_{no} + f_0(r) (1 - P_{no})} =\frac{e^{-\epsilon|r-1|}P_{no}}{ e^{-\epsilon|r-1|}P_{no} + e^{-\epsilon|r|}(1-P_{no})}
\end{equation*}
    
Similarly, the updated probability that the respondent did not answer NO is given by:
\begin{equation*}
    \frac{f_0(r) (1 - P_{no})}{f_1(r) P_{no} + f_0(r) (1 - P_{no})} =\frac{e^{-\epsilon|r|}(1-P_{no})}{ e^{-\epsilon|r-1|}P_{no} + e^{-\epsilon|r|}(1-P_{no})}
\end{equation*}

We can find the new threshold, $r_{\text{threshold}}$, by finding the value of $r$ for which the manager finds it equally likely that the respondent answered \no{} or did not answer \no{}. In other words, we set the above equations equal to each other and solve for $r$ to obtain:

$$r_{\text{threshold}} = \frac{\ln(1-P_{no}) - \ln(P_{no})}{2\epsilon} + \frac{1}{2}$$
when 
$\max\{\frac{1-P_{no}}{P_{no}}, \frac{P_{no}}{1-P_{no}}\} \leq e^{\epsilon}$ (i.e., as long as the prior is not too extreme).

Once this new threshold is obtained, it is straightforward to apply our explanation methods. We simply need to calculate the probability that the DP output will be greater than the threshold, given each choice the individual can make (e.g., participate or not participate). Note that in our study we use $0.5$ as the threshold. This is the value for $r_{\text{threshold}}$ that one obtains when $P_{no} = 0.5$.

\end{document}